# Predicting Multi-Order Magnetic Polariton Resonances for Radiative Properties Tailoring by Distributed Circuit Model

Hangjie Li[1], Junming Zhao[1,2*]

[1] *School of Energy Science and Engineering, Harbin Institute of Technology, Harbin 150001, China*

[2] *Key Laboratory of Aerospace Thermophysics, Ministry of Industry and Information Technology of China, Harbin 150001, China*

## Abstract

Surface plasmon polaritons (SPPs) and magnetic polaritons (MPs) are fundamental resonance modes that are widely used to tailor the thermal radiation properties of micro/nanostructured metamaterials. Lumped circuit models (LCMs) are usually constructed empirically to describe the MP resonance conditions, and different LCMs have to be constructed for different orders of MPs, but these are difficult to be built for high-order MP modes due to the complex electromagnetic field distribution. This work proposes a new type of circuit model, distributed circuit model (DCM), to describe and predict multi-order MP resonances inside the structure based on the minimum total impedance condition. This allows both fundamental and high-order MP resonances to be predicted with a unified circuit, significantly simplifying the analysis of high-order MPs. More importantly, the DCM shares a similar and clear physical picture as the LCM for describing MPs. The MP resonance conditions for four typical structures are derived. Theoretical predictions based on DCMs are compared with and validated by rigorous numerical simulations. This study deepens the understanding and facilitates the design of MP-based thermal radiation metamaterials.

* Corresponding author.

*Email address*: jmzhao@hit.edu.cn

# Highlights

- Proposed distributed circuit model to predict multi-order MP resonances
- Predict both fundamental and multi-order MP resonances by a unified circuit
- Derived multi-order MP resonance conditions for four typical structures
- Validated theoretical predictions by rigorous numerical simulations

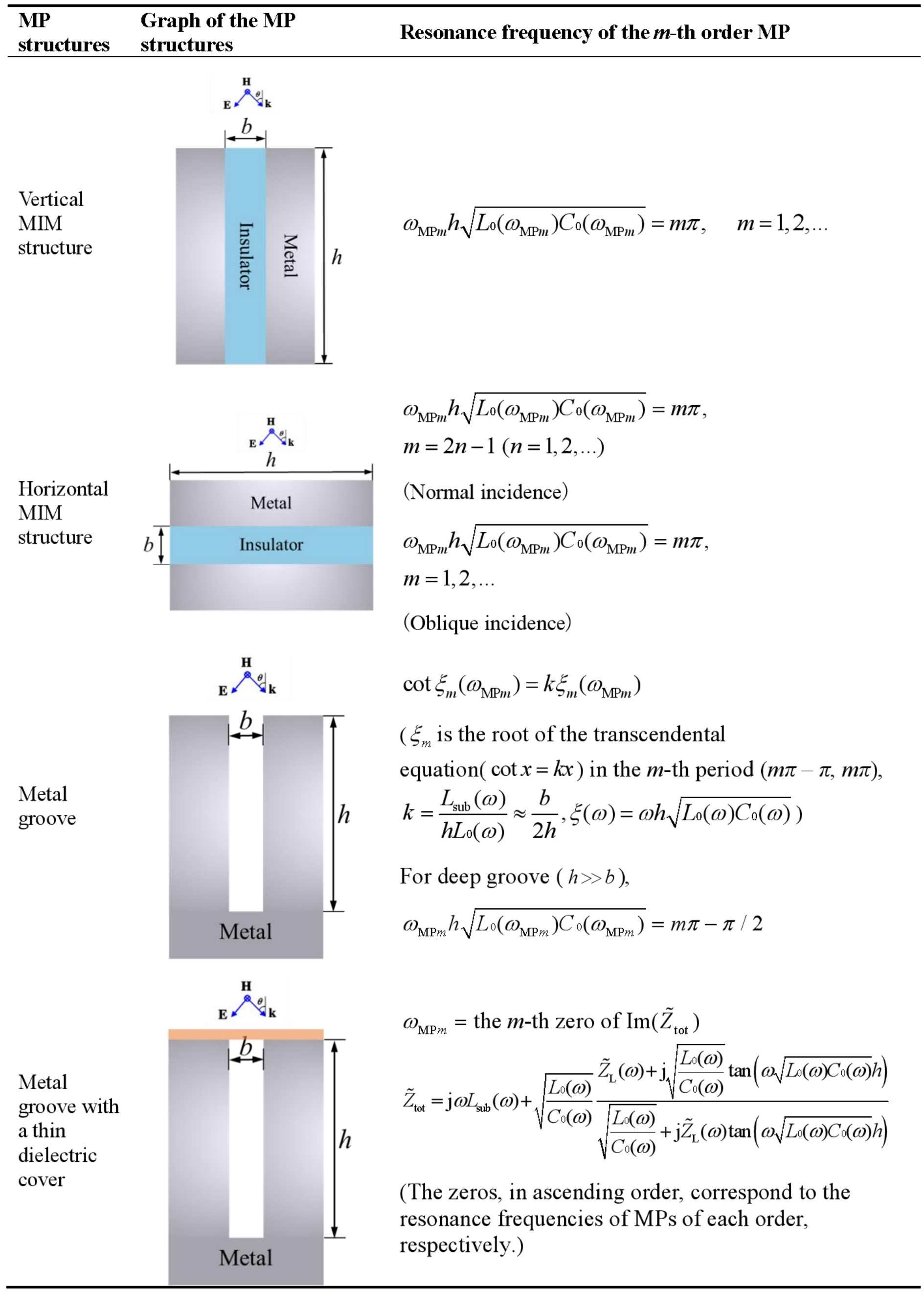

| MP structures | Graph of the MP structures | Resonance frequency of the *m*-th order MP |
|---|---|---|
| Vertical MIM structure | H, E, k, θ, b, h, Insulator, Metal | $\omega_{\mathrm{MP}m} h\sqrt{L_0(\omega_{\mathrm{MP}m})C_0(\omega_{\mathrm{MP}m})} = m\pi, \quad m = 1, 2, \ldots$ |
| Horizontal MIM structure | H, E, k, θ, h, b, Metal, Insulator | $\omega_{\mathrm{MP}m} h\sqrt{L_0(\omega_{\mathrm{MP}m})C_0(\omega_{\mathrm{MP}m})} = m\pi,$ $m = 2n-1\ (n = 1, 2, \ldots)$ (Normal incidence) $\omega_{\mathrm{MP}m} h\sqrt{L_0(\omega_{\mathrm{MP}m})C_0(\omega_{\mathrm{MP}m})} = m\pi,$ $m = 1, 2, \ldots$ (Oblique incidence) |
| Metal groove | H, E, k, θ, b, h, Metal | $\cot \xi_m(\omega_{\mathrm{MP}m}) = k\xi_m(\omega_{\mathrm{MP}m})$ ($\xi_m$ is the root of the transcendental equation( $\cot x = kx$ ) in the *m*-th period ($m\pi - \pi$, $m\pi$), $k = \frac{L_{\mathrm{sub}}(\omega)}{hL_0(\omega)} \approx \frac{b}{2h}, \xi(\omega) = \omega h\sqrt{L_0(\omega)C_0(\omega)}$ ) For deep groove ( $h >> b$ ), $\omega_{\mathrm{MP}m} h\sqrt{L_0(\omega_{\mathrm{MP}m})C_0(\omega_{\mathrm{MP}m})} = m\pi - \pi/2$ |
| Metal groove with a thin dielectric cover | H, E, k, θ, b, h, Metal | $\omega_{\mathrm{MP}m}$ = the *m*-th zero of $\mathrm{Im}(\tilde{Z}_{\mathrm{tot}})$ $\tilde{Z}_{\mathrm{tot}} = \mathrm{j}\omega L_{\mathrm{sub}}(\omega) + \sqrt{\frac{L_0(\omega)}{C_0(\omega)}} \frac{\tilde{Z}_{\mathrm{L}}(\omega) + \mathrm{j}\sqrt{\frac{L_0(\omega)}{C_0(\omega)}} \tan\left(\omega\sqrt{L_0(\omega)C_0(\omega)}h\right)}{\sqrt{\frac{L_0(\omega)}{C_0(\omega)}} + \mathrm{j}\tilde{Z}_{\mathrm{L}}(\omega)\tan\left(\omega\sqrt{L_0(\omega)C_0(\omega)}h\right)}$ (The zeros, in ascending order, correspond to the resonance frequencies of MPs of each order, respectively.) |

## Nomenclature

| | |
|---|---|
| $b$ | slit width, m |
| $\tilde{\mathbf{B}}$ | magnetic flux density vector, Wb/m$^2$ |
| $C$ | capacitance, F |
| $C_0$ | shunt capacitance per unit length in a distributed circuit, F/m |
| $d$ | length of a control volume, m |
| $\tilde{\mathbf{D}}$ | electric displacement field vector, C/m$^2$ |
| $\tilde{\mathbf{E}}$ | electric field intensity vector, V/m |
| $h$ | slit length, m |
| $\tilde{\mathbf{H}}$ | magnetic field intensity vector, A/m |
| $\tilde{I}$ | current, A |
| $\tilde{I}_{\mathrm{L}}$ | distributed circuit terminal current, A |
| $J$ | dimensionless current density |
| $\tilde{\mathbf{J}}$ | total current density, A/m$^2$ |
| $\tilde{\mathbf{J}}_{\mathrm{f}}$ | current density of free charges, A/m$^2$ |
| $k_0$ | vacuum wave number, rad/m |
| $k$ | defined by Eq. (29) |
| $L$ | inductance, H |
| $L_0$ | series inductance per unit length in a distributed circuit, H/m |
| $m$ | the order of the resonance mode |
| $n_{\mathrm{GP}}$ | the real part of the effective mode index of the GP |
| $R$ | resistance, Ω |
| $S$ | cross section area of a control volume, m$^2$ |
| $\tilde{U}$ | voltage, V |
| $\tilde{U}_{\mathrm{L}}$ | distributed circuit terminal voltage, V |
| $X$ | reactance, Ω |
| $\tilde{Z}$ | impedance, Ω |

$\tilde{Z}_{\mathrm{L}}$ load impedance, Ω

$\tilde{Z}_{\mathrm{sub}}$ substrate impedance, Ω

$\tilde{Z}_{\mathrm{in,\ load}}$ input impedance at the initial end when the terminal is connected to a load, Ω

$\tilde{Z}_{\mathrm{in,\ open}}$ input impedance at the initial end when the terminal is open-circuited, Ω

$\gamma$ the phase constant, rad/m

$\Gamma$ scattering rate, rad/s

$\delta$ the equivalent current depth, m

$\tilde{\varepsilon}$ complex dielectric function

$\varepsilon'$ real part of complex dielectric function

$\varepsilon''$ imaginary part of complex dielectric function

$\varepsilon_0$ vacuum permittivity, $8.8542 \times 10^{-12}$ F/m

$\varepsilon_\infty$ high-frequency dielectric function

$\theta$ incidence angle, °

$\kappa$ extinction coefficient

$\mu_0$ vacuum permeability, $4\pi\times10^{-7}$ H/m

$\xi$ defined by Eq. (29)

$\rho_{\mathrm{f}}$ free charge density, C/m$^3$

$\rho_R$ resistivity, Ω·m

$\rho_L$ the inductance with unit length and unit cross-sectional area defined by Eq. (8), H·m

$\rho_C$ the capacitance with unit length and unit cross-sectional area defined by Eq. (8), F/m

$\tilde{\sigma}$ complex conductivity, A/(F·m)

$\phi$ reflection phase

$\omega$ angular frequency, rad/s

$\omega_p$ plasma frequency, rad/s

$\omega_{\mathrm{MP}m}$ resonance frequency of the $m$-th order MP, rad/s

# 1. Introduction

Metamaterials provide a promising platform for tailoring thermal radiation properties and have been widely utilized in various infrared applications, including solar energy harvesting [1-3], thermophotovoltaic devices [4-7], radiative cooling [8-11], gas and biochemical sensing [12-14], and thermal camouflage [15-20]. Among these applications, magnetic polaritons (MPs) are one of the most significant resonance modes for controlling the spectral thermal radiation properties of metamaterials. They refer to the coupling between incident electromagnetic waves and magnetic excitations within the micro/nanostructured metamaterials [4, 9, 19, 21].

The lumped circuit model (LCM) is a reliable method to describe and predict the MP resonance conditions in micro/nanostructures [22-25], as shown in **Fig. 1** (a). The general idea is to treat the microstructures as various "lumped" elements, such as inductors or capacitors. The resonant frequency is determined by minimizing the modulus of the total complex impedance of the circuit $\omega_{\mathrm{R}} = \arg\min\left|\tilde{Z}(\omega)\right|$, which is typically accomplished by setting the imaginary part of the impedance to zero [23].

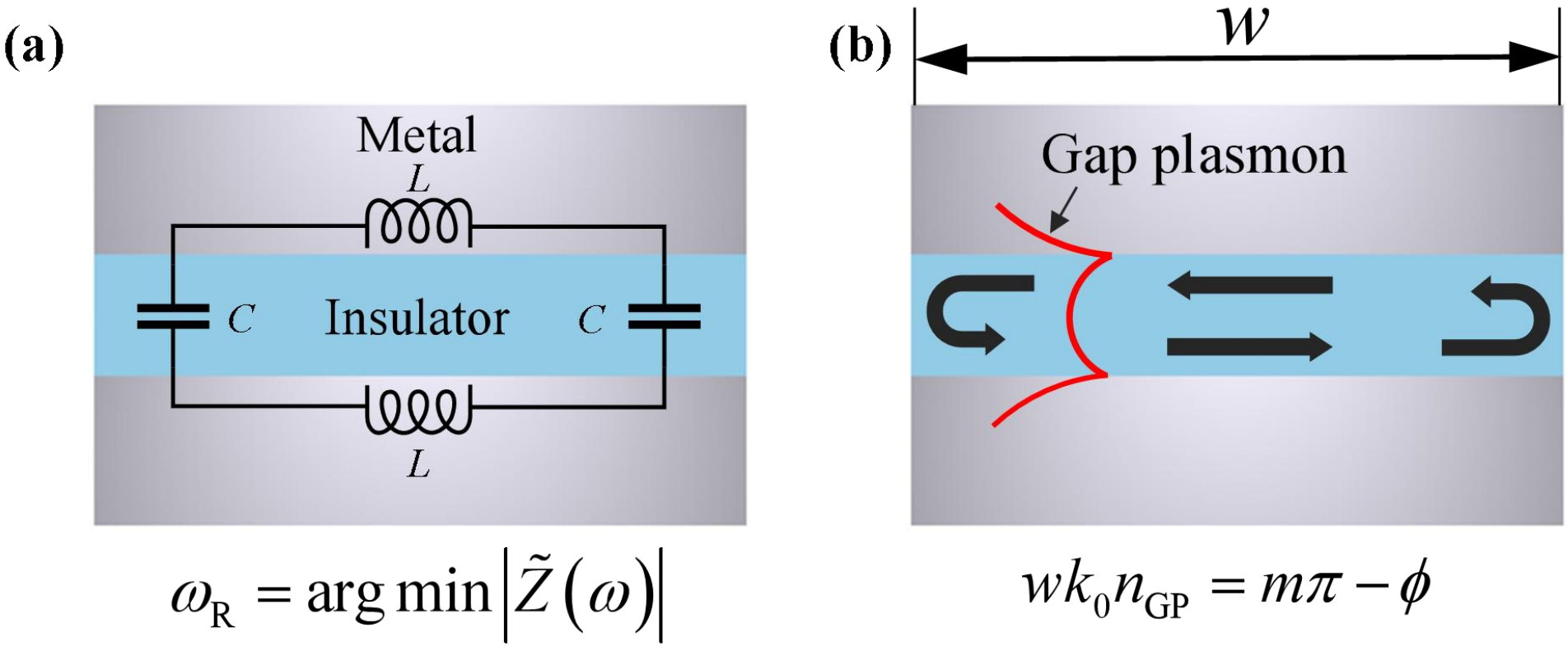

**Fig. 1** Schematic diagrams of two viewpoints for understanding the resonance within the MIM structure. **(a)** A schematic diagram of the LCM applied for MPs in the MIM structure. **(b)** A schematic diagram of the Fabry-Perot resonator model for GPs in the MIM structure.

Subsequently, a number of improved LCMs were employed. Sakurai et al. [26] developed a resistor–inductor–capacitor (RLC) model, which can not only predict the MP resonance frequency but also the full width at half maximum and quality factor. Li et al. [27] enhanced the predictive accuracy of the LCM by considering boundary effects and introducing a parasitic capacitance. Qing et al. [28] studied the coupling behavior of graphene surface plasmons (GSP) and MP by using the coupled oscillator model and obtained the dispersion relation of the hybrid mode. Long et al. [29] enhanced the absorption of monolayer $MoS_2$ by exciting multi-order MPs within a silver grating. They further extended the LCM to a generalized one for predicting the resonance wavelengths of multi-order MPs. Guo et al. [30, 31] proposed a Mutual inductor-inductor-capacitor (MLC) circuit, treating the interaction between surface plasmon polaritons (SPPs)/surface phonon polaritons (SPhPs) and MPs as mutual inductance to reduce the deviation. Furthermore, they proposed a multi-mode coupled LC (MCLC) circuit model to describe and predict the resonance conditions of multi-mode MPs. Multi-mode and multi-order MP resonances are regarded as the superposition of the resonances of each current loop. The reciprocal of the overall resonance frequency is equal to the geometric mean of the reciprocals of the resonance frequencies of each current loop generated by MPs [32]. However, when predicting high-order MP resonance conditions, these improved LCMs have to construct new circuits for different orders of MPs. Moreover, building such models becomes increasingly difficult for higher-order MP modes due to the complex electromagnetic field distribution.

When using the traditional LCM to predict high-order MP resonances, it is necessary to construct multiple independent current loops. This requirement can only be satisfied when the lengths of these loops match the dimensions determined through electromagnetic field distribution analysis [29, 30]. At present, it remains challenging to simultaneously predict both fundamental and high-order MP resonances using a single, unified circuit model.

In the Metal-Insulator-Metal (MIM) structure, some researchers attribute the scattering resonances and resonant fields, which are similar to those of MPs, to GPs [33-36], as shown in **Fig. 1** (b). The GPs are formed by the coupling of surface plasmon polaritons (SPPs) at the two metal-dielectric interfaces [37, 38]. In a MIM structure with finite width, the GP mode can undergo multiple reflections, leading to the formation of a standing wave GP mode. The resonant position of this mode can be accurately described using the classical Fabry-Perot resonator model $wk_0n_{\mathrm{GP}} = m\pi - \phi$ [33, 39], where $w$ is the width of the MIM structure, $k_0$ is the vacuum wave number, $n_{\mathrm{GP}}$ is the real part of the effective mode index of the GP, $m$ is an integer defining the order of the mode, and $\phi$ is an additional phase shift acquired following GP reflection at the boundary of the MIM structure. Some studies have shown that in MIM structures, the resonant frequency indicated by the Fabry-Perot resonator model of GPs exhibits similarity to the resonant frequency of the LCM associated with MPs [40], indicating the equivalence of the MP and GP to some extent. Recently, Gong et al. [41] proposed using the propagation constant of the transmission line model to establish the resonance conditions of MPs based on the physical picture of standing-wave, i.e., Fabry-Perot resonator model or gap plasmons (GPs), and the phase shift $\phi$ was determined empirically. Since the physical pictures of the two models (i.e., circuit model and Fabry-Perot resonator model) and the resonance conditions are much different, it is difficult to claim that the MPs and GPs are exactly the same. More importantly, it should be noted that $\phi$ is a fitting parameter, and the fitting process must be performed independently for each order of the GP mode. In general, theoretical prediction of the resonance conditions of high-order MPs remains challenging. There is an urgent need to develop a unified theoretical framework capable of simultaneously predicting both fundamental and high-order MPs for different structures.

In this work, a new type of circuit model, distributed circuit model (DCM), is proposed to describe and predict multi-order MP resonances inside the structure. The greatest advantage of the DCM is its ability to predict both fundamental and high-order MP resonances with a unified circuit, significantly simplifying the analysis of high-order MPs. We integrate the transmission line impedance with the physical properties and resonance mechanism of MPs. Based on this, a more stringent set of resonance conditions has been systematically deduced. Corresponding DCMs are established for four typical structures and the MP resonance conditions are derived. A detailed analysis of the distribution of electromagnetic fields and currents within these structures under resonant conditions elucidates the rationale for the DCM. Theoretical predictions based on DCMs are compared with and validated by the calculation results of rigorous-coupled wave analysis (RCWA).

## 2. Theoretical Model

### 2.1 General definition of the circuit impedance

Before formally introducing the DCM, we first briefly explain how to analyze the circuit impedance starting from the dielectric function. If the field variables are time harmonics with time dependency of $e^{j\omega t}$, the Maxwell equations can be rewritten in frequency domain as [42]

$$\nabla \times \tilde{\mathbf{H}} = \tilde{\mathbf{J}}_{\mathrm{f}} + \mathrm{j}\omega\tilde{\mathbf{D}} = \tilde{\mathbf{J}} \tag{1}$$

$$\nabla \times \tilde{\mathbf{E}} = -\mathrm{j}\omega\tilde{\mathbf{B}} \tag{2}$$

$$\nabla \bullet \tilde{\mathbf{D}} = \rho_{\mathrm{f}} \tag{3}$$

$$\nabla \bullet \tilde{\mathbf{B}} = 0 \tag{4}$$

where $\mathrm{j} = \sqrt{-1}$, $\omega$ is the angular frequency, $\tilde{\mathbf{E}}$ and $\tilde{\mathbf{H}}$ are the electric and magnetic field intensity vector respectively, $\tilde{\mathbf{J}}_{\mathrm{f}}$ is current density of free charges, $\tilde{\mathbf{J}}$ is the total current density, $\tilde{\mathbf{D}}$ is the electric displacement field vector, $\tilde{\mathbf{B}}$ is the magnetic flux density vector, $\rho_{\mathrm{f}}$ is the free charge density.

A circuit can be defined for a control volume of media based on the Ohm's law. The general Ohm's law for total current density $\tilde{\mathbf{J}}$ including the displacement current can be written as [42]

$$\tilde{\mathbf{J}} = \tilde{\sigma}(\omega)\tilde{\mathbf{E}} \tag{5}$$

There is a relationship between the complex conductivity $\tilde{\sigma}$ and the complex dielectric function $\tilde{\varepsilon}$ given by $\tilde{\sigma}(\omega) = \mathrm{j}\omega\varepsilon_0\tilde{\varepsilon}(\omega) = \omega\varepsilon_0\varepsilon''(\omega) + \mathrm{j}\omega\varepsilon_0\varepsilon'(\omega)$ [43], where $\varepsilon_0$ is the vacuum permittivity and $\tilde{\varepsilon}(\omega) = \varepsilon'(\omega) - \mathrm{j}\varepsilon''(\omega)$ (the negative sign of the imaginary part is caused by the time phase factor $e^{j\omega t}$ of the wave function). Considering a cube of length $d$ and cross section area of $S$ as depicted in **Fig. 2**(a) as a control volume, Eq. (5) can be rewritten as

$$\tilde{\mathbf{J}}S = \left[\tilde{\sigma}(\omega)\frac{S}{d}\right]\left[\tilde{\mathbf{E}}d\right] \tag{6}$$

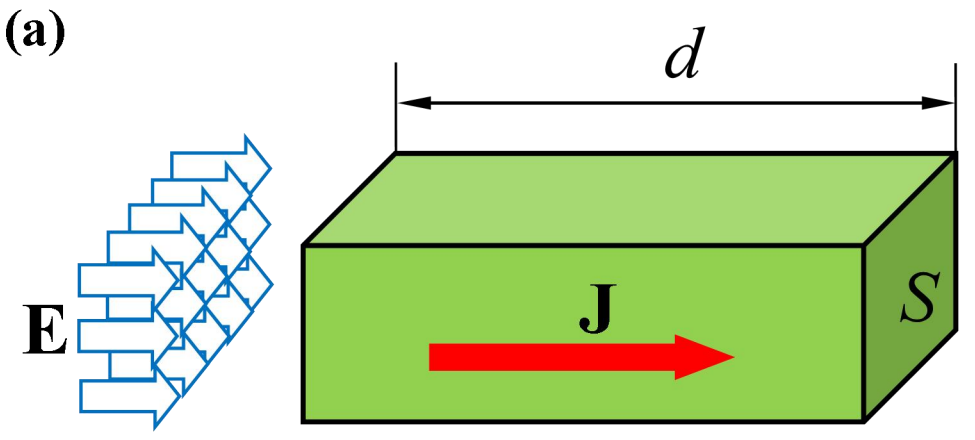

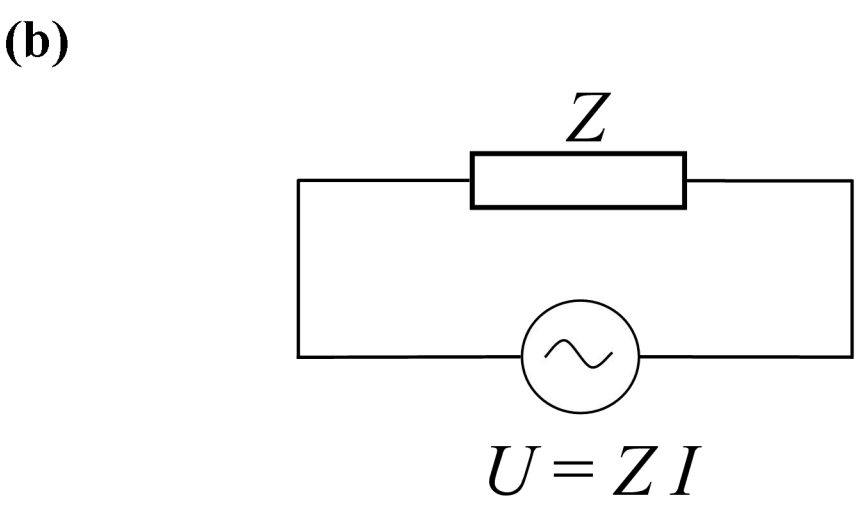

**Fig. 2** General equivalent circuit for a control volume. **(a)** Geometry of the selected control volume, and **(b)** the equivalent circuit.

Assume the direction of electric field is along the length of cube and the electric field and current density are uniform, then the electric current and voltage can be defined as $\tilde{I} = \tilde{J}S$ and $\tilde{U} = \tilde{E}d$ (note: the current density and electric field have the same direction, so we use scalar form for brevity), thus an equivalent circuit can be built as shown in **Fig. 2**(b).

$$\tilde{U} = \tilde{Z}(\omega)\tilde{I} \tag{7}$$

where the impedance $\tilde{Z}(\omega) = \tilde{\sigma}(\omega)^{-1}\frac{d}{S}$. This is a general circuit and the impedance is defined by the complex conductivity (or permittivity), which does not give the micro structure of the media. The explicit expression of $\tilde{Z}(\omega)$ can be formulated as

$$\begin{aligned}\tilde{Z}(\omega) &= \frac{\omega\varepsilon_0\varepsilon''(\omega)}{\left(\omega\varepsilon_0\varepsilon''(\omega)\right)^2 + \left(\omega\varepsilon_0\varepsilon'(\omega)\right)^2}\frac{d}{S} + \frac{-\mathrm{j}\omega\varepsilon_0\varepsilon'(\omega)}{\left(\omega\varepsilon_0\varepsilon''(\omega)\right)^2 + \left(\omega\varepsilon_0\varepsilon'(\omega)\right)^2}\frac{d}{S} \\ &= \begin{cases} R(\omega) + \mathrm{j}\omega L(\omega) = \rho_R(\omega)\dfrac{d}{S} + \mathrm{j}\omega\rho_L(\omega)\dfrac{d}{S} & \left(\varepsilon'(\omega) \le 0\right) \\ R(\omega) + \dfrac{1}{\mathrm{j}\omega C(\omega)} = \rho_R(\omega)\dfrac{d}{S} + \dfrac{1}{\mathrm{j}\omega\rho_C(\omega)}\dfrac{d}{S} & \left(\varepsilon'(\omega) \ge 0\right) \end{cases}\end{aligned} \tag{8}$$

where

$$\rho_R(\omega) = \frac{\omega\varepsilon_0\varepsilon''(\omega)}{\left(\omega\varepsilon_0\varepsilon''(\omega)\right)^2 + \left(\omega\varepsilon_0\varepsilon'(\omega)\right)^2} \tag{9}$$

$$\rho_L(\omega) = \frac{-\varepsilon'(\omega)}{\omega^2\varepsilon_0\left(\varepsilon''(\omega)^2 + \varepsilon'(\omega)^2\right)} \tag{10}$$

$$\rho_C(\omega) = \varepsilon_0 \frac{\left(\varepsilon''(\omega)^2 + \varepsilon'(\omega)^2\right)}{\varepsilon'(\omega)} \tag{11}$$

In which $\rho_R$, $\rho_L$, and $\rho_C$ denote the resistance, inductance, and capacitance per unit length and per unit cross-sectional area, respectively. This indicates that when the material is a conventional dielectric (such as $SiO_2$ or Si), with its real part of the dielectric function $\varepsilon' > 0$, it exhibits capacitive impedance. In contrast, if the material possesses a negative real part of the dielectric function ($\varepsilon' < 0$) at optical wavelengths (such as noble metals like Ag and Au), it exhibits inductive impedance. Note that the equivalent inductance $L$ here is only the kinetic inductance.

### 2.2 Input impedance of the distributed circuit

For DCM, we still derive the resonant frequency starting from the circuit impedance. We derive the circuit impedance through the distributed circuit and further determine the specific resonance conditions in combination with the MP resonance condition ( $\omega_{\mathrm{R}} = \arg\min\left|\tilde{Z}(\omega)\right|$ ). In this approach, it is a key element to obtain the input impedance of the basic MIM structure, which will be discussed in detail in the following.

Take the equivalent circuit at the typical MIM slit as an instance. It is assumed that the left side is the initial end and the right side is the terminal end. When the directions of the electric fields at all points between the metal walls are the same, the DCM is depicted as shown in **Fig. 3** (a). The entire distributed circuit consists of an infinite number of small line elements, where $R_0$ denotes the series resistance per unit length, $L_0$ represents the series inductance per unit length, $G_0$ indicates the shunt conductance per unit length, and $C_0$ stands for the shunt capacitance per unit length. Here, the distribution parameters are assumed to be uniformly distributed along the $z$ direction. The wall current gradually decreases from the starting end to the terminal end. For metals, the real part of their dielectric function is typically smaller than 0 and can be equivalently considered as an inductive impedance. For insulators, the real part of their dielectric function is generally larger than 0 and can be equivalently considered as a capacitive impedance. The fact that the electric field is present everywhere between the two metal walls when the MP resonance is excited implies that the displacement current branches at all locations on the metal walls. Hence, it is more reasonable to describe and predict MP resonances using the DCM rather than the LCM.

(a)

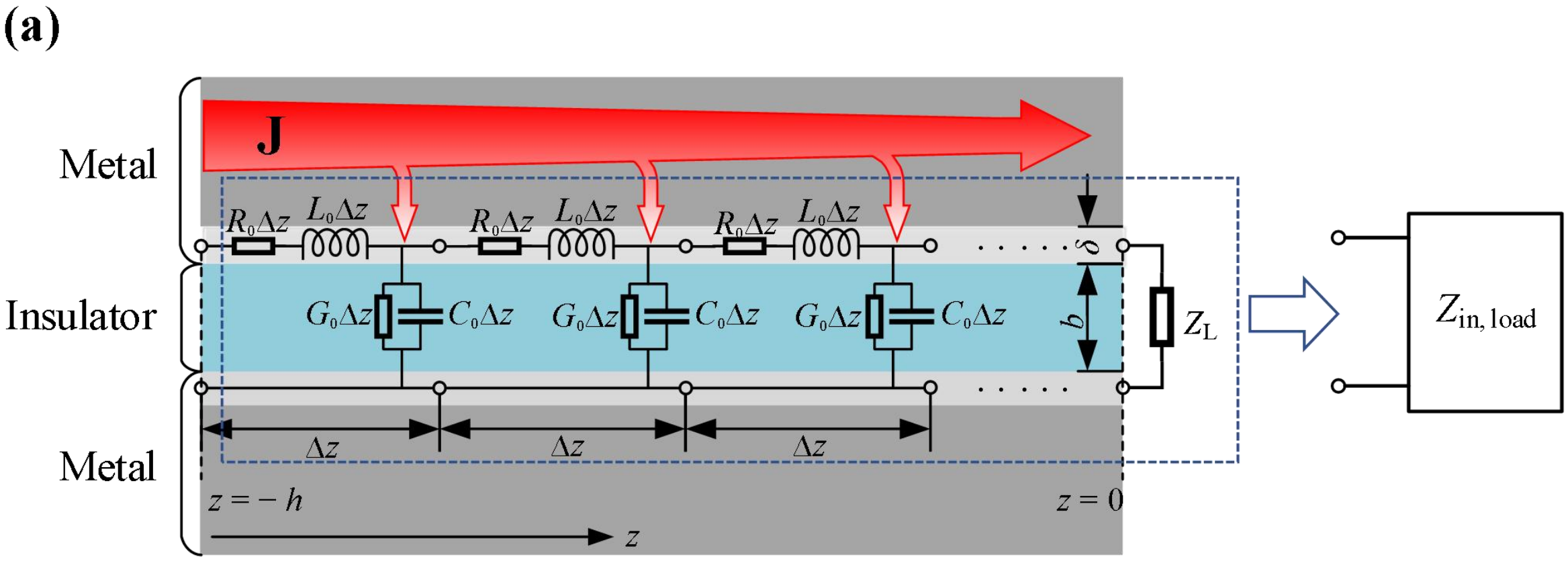

(b)

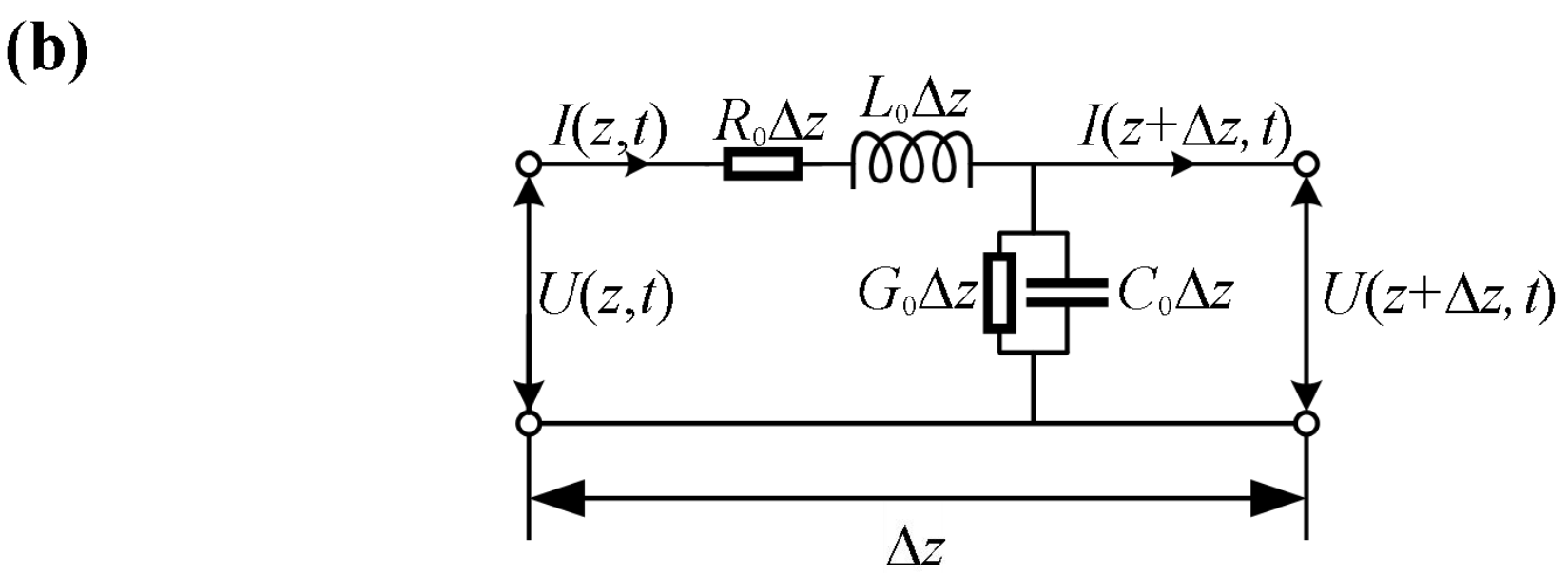

**Fig. 3** Schematic diagram of the DCM. **(a)** A DCM and **(b)** a line element of length Δz in the DCM. $R_0$, $L_0$, $G_0$ and $C_0$ are the distribution parameters in the DCM.

Now, we redraw the equivalent circuit of the small line element from $z$ to $z+\Delta z$ as shown in **Fig. 3** (b). Let $U(z,t)$ and $I(z,t)$ represent the voltage and current at $z$, and $U(z+\Delta z,t)$ and $I(z+\Delta z,t)$ denote the voltage and current at $(z+\Delta z)$. The analysis of the input impedance is aligned with the principles of transmission line theory. For the infinitesimal segment shown in **Fig. 3** (b), according to Kirchhoff's voltage and current laws, the corresponding differential equations in the frequency domain is [44]

$$\frac{d\tilde{U}(z)}{dz} = -\left(R_0(\omega)+\mathrm{j}\omega L_0(\omega)\right)\tilde{I}(z) \tag{12}$$

$$\frac{d\tilde{I}(z)}{dz} = -\left(G_0(\omega)+\mathrm{j}\omega C_0(\omega)\right)\tilde{U}(z) \tag{13}$$

Thereby, the expression of the input impedance at the initial end can be derived,

$$\tilde{Z}_{\text{in, load}}\Big|_{z=-h} = \frac{\tilde{U}(-h)}{\tilde{I}(-h)} = \tilde{Z}_0 \frac{\tilde{Z}_L + \tilde{Z}_0 \tanh\left(\sqrt{\left(R_0(\omega)+\mathrm{j}\omega L_0(\omega)\right)\left(G_0(\omega)+\mathrm{j}\omega C_0(\omega)\right)}h\right)}{\tilde{Z}_0 + \tilde{Z}_L \tanh\left(\sqrt{\left(R_0(\omega)+\mathrm{j}\omega L_0(\omega)\right)\left(G_0(\omega)+\mathrm{j}\omega C_0(\omega)\right)}h\right)} \tag{14}$$

Where, $\tilde{Z}_L = \tilde{U}(0)/\tilde{I}(0)$ is the load impedance, $\tilde{Z}_0 = \sqrt{\frac{R_0(\omega)+\mathrm{j}\omega L_0(\omega)}{G_0(\omega)+\mathrm{j}\omega C_0(\omega)}}$ is the characteristic impedance of the transmission line. In the case of an open-circuited terminal, $\tilde{Z}_L \to \infty$, the input impedance at the initial end can be simplified to

$$\tilde{Z}_{\text{in, open}}\big|_{z=-h} = \frac{\tilde{Z}_0}{\tanh\left(\sqrt{\left(R_0(\omega)+\mathrm{j}\,\omega L_0(\omega)\right)\left(G_0(\omega)+\mathrm{j}\,\omega C_0(\omega)\right)}h\right)} \tag{15}$$

Considering both the slit width $b$ and the thickness $l$ in the direction perpendicular to the paper surface (which can be set to unity for a one-dimensional grating), $R_0$, $G_0$, $C_0$ and $L_0$ can be derived from Eq. (8) as follows

$$R_0(\omega) = \rho_{R,\mathrm{m}}(\omega)\frac{2}{\delta(\omega)l} \tag{16}$$

$$G_0(\omega) = \frac{\rho_{R,\mathrm{i}}(\omega)\frac{b}{l}}{\left|\rho_{R,\mathrm{i}}(\omega)\frac{b}{l} + \frac{1}{\mathrm{j}\omega\rho_{C,\mathrm{i}}(\omega)}\frac{b}{l}\right|^2} \tag{17}$$

$$C_0(\omega) = \rho_{C,\mathrm{i}}(\omega)\frac{l}{b} \tag{18}$$

$$L_0(\omega) = L_{\mathrm{K}0}(\omega) + L_{\mathrm{M}0} = \rho_{L,\mathrm{m}}(\omega)\frac{2}{\delta(\omega)l} + \frac{\mu_0 b}{l} \tag{19}$$

Here, $L_0$ is composed of the kinetic inductance $L_{\mathrm{K}0}$ and mutual inductance $L_{\mathrm{M}0}$; the coefficient 2 in $R_0$ and $L_{\mathrm{K}0}$ implies that the resistance and kinetic inductance are formed by two metallic segments—upper and lower—connected in series. $\delta = c_1\lambda/\left(2\pi\kappa\right)$ is the equivalent current depth, in which a correction coefficient $c_1$ is included to address the non-uniformly distribution of the electric current [26, 32] (Note that the penetration depth of the electric field is $\delta = \lambda/\left(2\pi\kappa\right)$, where $\kappa$ is the extinction coefficient). In practice, the correction coefficient $c_1$ is fitted. $\mu_0$ is the vacuum permeability. $\rho_{R,\mathrm{m}}(\omega)$, $\rho_{L,\mathrm{m}}(\omega)$, $\rho_{R,\mathrm{i}}$ and $\rho_{C,\mathrm{i}}(\omega)$ can be obtained from Eqs. (9), (10) and (11), the subscripts 'm' and 'i' respectively denote metal and insulator.

### 2.3 DCM for typical MP structures

In the discussion in Section 2.2, we treat the entire slit as a circuit, where the direction of the electric field is the same at all internal points for processing. However, in certain structures, when MPs are excited, electric fields in different directions are present within the metal slit. The resonance condition cannot be simply expressed using a transmission line model; it requires special treatment based on the specific circuit characteristics of different structures. **Fig. 4** shows the DCMs for four typical MP structures.

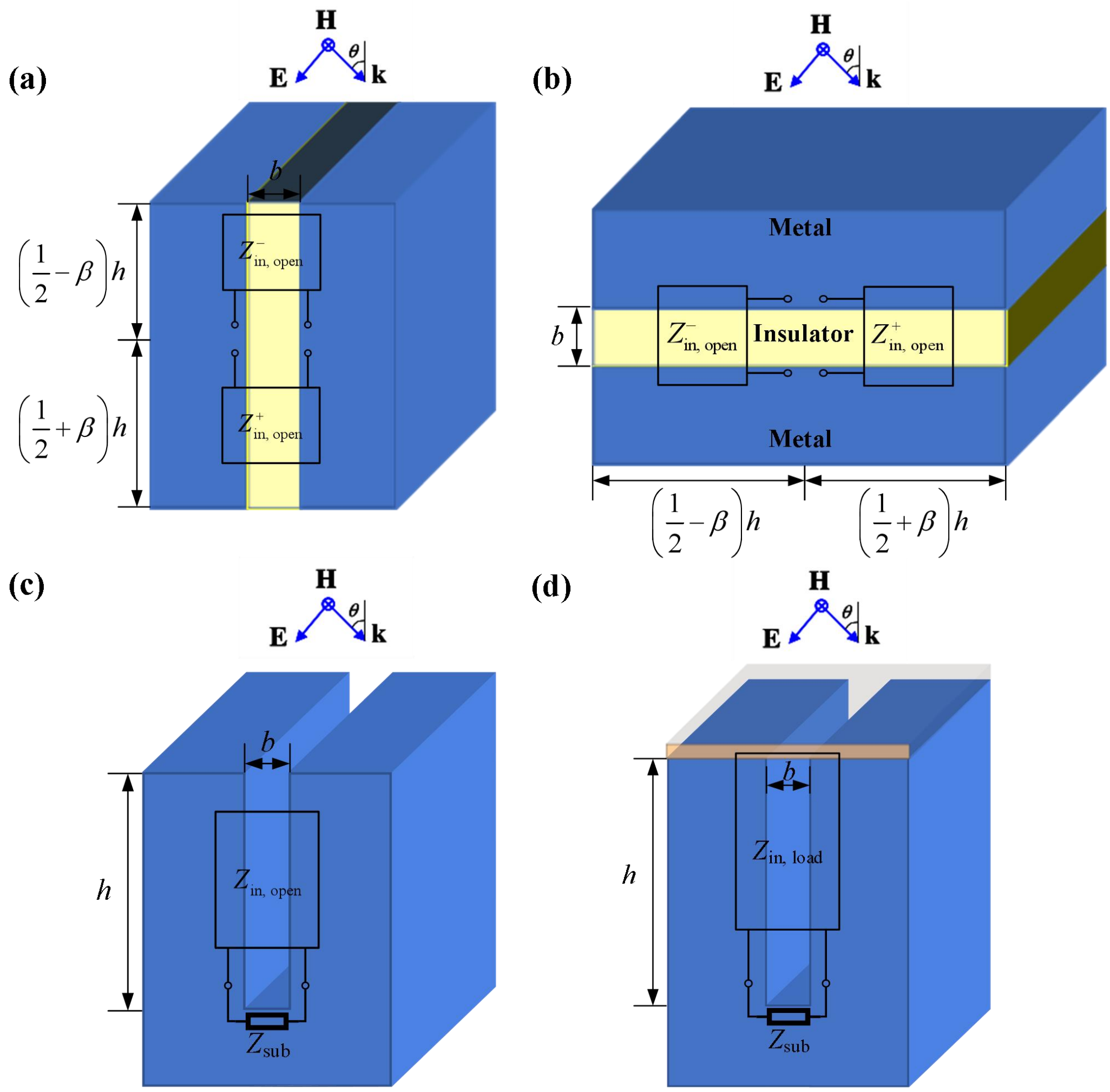

**Fig. 4** DCMs for typical structures that supporting multi-order MPs. **(a)** vertical MIM structure; **(b)** horizontal MIM structure; **(c)** metal groove; **(d)** metal groove with a thin dielectric layer.

For the vertical MIM structure, we divide the slit into two unequal parts, one with a length of $\left(\frac{1}{2}+\beta\right)h$ and the other with a length of $\left(\frac{1}{2}-\beta\right)h$, as depicted in **Fig. 4** (a). The initial end of the distributed circuit is located inside the slit. A detailed explanation will be discussed in the next section. The upper and lower parts are in series. The total input impedance is the sum of the input impedances of the upper and lower parts,

$$\begin{aligned}\tilde{Z}_{\text{tot}} &= \tilde{Z}_{\text{in, open}}\Big|_{z=-\left(\frac{1}{2}+\beta\right)h} + \tilde{Z}_{\text{in, open}}\Big|_{z=-\left(\frac{1}{2}-\beta\right)h} \\ &= \frac{\tilde{Z}_0}{\tanh\left(\sqrt{\left(R_0(\omega)+\mathrm{j}\omega L_0(\omega)\right)\left(G_0(\omega)+\mathrm{j}\omega C_0(\omega)\right)}\left(\frac{1}{2}+\beta\right)h\right)} \\ &+ \frac{\tilde{Z}_0}{\tanh\left(\sqrt{\left(R_0(\omega)+\mathrm{j}\omega L_0(\omega)\right)\left(G_0(\omega)+\mathrm{j}\omega C_0(\omega)\right)}\left(\frac{1}{2}-\beta\right)h\right)}\end{aligned} \tag{20}$$

The specific resonance frequency can now be determined by incorporating the MP resonance condition ( $\omega_{\mathrm{R}} = \arg\min\left|\tilde{Z}_{\mathrm{tot}}(\omega)\right|$ ). However, to derive a concise expression for the resonance condition, a lossless approximation ($R_0$ = 0 and $G_0$ = 0) is introduced, and its validity will be discussed subsequently. Under the lossless approximation, Eqs. (14) and (15) can be simplified to

$$\tilde{Z}_{\text{in, load}}\Big|_{z=-h} = \frac{\tilde{U}(-h)}{\tilde{I}(-h)} = \sqrt{\frac{L_0(\omega)}{C_0(\omega)}} \frac{\tilde{Z}_{\mathrm{L}} + \mathrm{j}\sqrt{\frac{L_0(\omega)}{C_0(\omega)}}\tan\left(\omega\sqrt{L_0(\omega)C_0(\omega)}h\right)}{\sqrt{\frac{L_0(\omega)}{C_0(\omega)}} + \mathrm{j}\tilde{Z}_{\mathrm{L}}\tan\left(\omega\sqrt{L_0(\omega)C_0(\omega)}h\right)} \tag{21}$$

$$\tilde{Z}_{\text{in, open}}\Big|_{z=-h} = -\mathrm{j}\sqrt{\frac{L_0(\omega)}{C_0(\omega)}}\cot\left(\omega\sqrt{L_0(\omega)C_0(\omega)}h\right) \tag{22}$$

Based on this approximation, $\tilde{Z}_{\mathrm{tot}}$ can be simplified to

$$\begin{aligned}\tilde{Z}_{\mathrm{tot}} &= \tilde{Z}_{\text{in, open}}\Big|_{z=-\left(\frac{1}{2}+\beta\right)h} + \tilde{Z}_{\text{in, open}}\Big|_{z=-\left(\frac{1}{2}-\beta\right)h} \\ &= -\mathrm{j}\sqrt{\frac{L_0(\omega)}{C_0(\omega)}}\left\{\cot\left[\left(\frac{1}{2}+\beta\right)h\omega\sqrt{L_0(\omega)C_0(\omega)}\right] + \cot\left[\left(\frac{1}{2}-\beta\right)h\omega\sqrt{L_0(\omega)C_0(\omega)}\right]\right\} \\ &= -\mathrm{j}\sqrt{\frac{L_0(\omega)}{C_0(\omega)}}\left\{\frac{2\tan\left(\frac{1}{2}\omega h\sqrt{L_0(\omega)C_0(\omega)}\right)\left[1+\tan^2\left(\beta\omega h\sqrt{L_0(\omega)C_0(\omega)}\right)\right]}{\tan^2\left(\frac{1}{2}\omega h\sqrt{L_0(\omega)C_0(\omega)}\right) - \tan^2\left(\beta\omega h\sqrt{L_0(\omega)C_0(\omega)}\right)}\right\}\end{aligned} \tag{23}$$

The impedance is approximately a pure imaginary number. When $\tan\left(\frac{1}{2}\omega h\sqrt{L_0(\omega)C_0(\omega)}\right) = 0$ or $\tan^2\left(\frac{1}{2}\omega h\sqrt{L_0(\omega)C_0(\omega)}\right) - \tan^2\left(\beta\omega h\sqrt{L_0(\omega)C_0(\omega)}\right) = \infty$ , $\tilde{Z}_{\mathrm{tot}} \to 0$ . $\tan\left(\frac{1}{2}\omega h\sqrt{L_0(\omega)C_0(\omega)}\right) = 0$ is equivalent to

$$\omega_{\mathrm{MP}m} h\sqrt{L_0(\omega_{\mathrm{MP}m})C_0(\omega_{\mathrm{MP}m})} = m\pi, \qquad m = 2n\ (n = 1, 2, \ldots) \tag{24}$$

corresponding to the resonance condition of even-order MP resonances. $\tan^2\left(\frac{1}{2}\omega h\sqrt{L_0(\omega)C_0(\omega)}\right) - \tan^2\left(\beta\omega h\sqrt{L_0(\omega)C_0(\omega)}\right) = \infty$ is equivalent to

$$\omega_{\mathrm{MP}m} h\sqrt{L_0(\omega_{\mathrm{MP}m})C_0(\omega_{\mathrm{MP}m})} = m\pi, \qquad m = 2n-1\ (n = 1, 2, \ldots) \tag{25}$$

corresponding to the resonance condition of odd-order MP resonances. Therefore, the resonance condition of the metal grating without a metal substrate can be obtained as

$$\omega_{\mathrm{MP}m} h\sqrt{L_0(\omega_{\mathrm{MP}m})C_0(\omega_{\mathrm{MP}m})} = m\pi, \qquad m = 1, 2, \ldots \tag{26}$$

where *m* is the resonance order of MPs. It can be seen that the resonance condition is independent of the selection of *β*. When the slit is symmetrically divided (*β* = 0), only resonant conditions for odd-order MP resonances can be obtained. The resonance condition will be validated in the next section. According to Eq. (26), in the absence of no dispersion (i.e., when $L_0$ and $C_0$ are frequency-independent), the MP resonance frequencies predicted by the equation are equally spaced.

The validity of the lossless approximation can be expounded upon from the perspective of the LCM. As is well known, when MP*m* is excited, *m* current loops are correspondingly formed within the structure, as depicted in **Fig. 5**. Each loop can be associated with an individual lumped circuit of the same resonance frequency. In a single current loop, the resistance, inductor and capacitor are connected in series. In an RLC series circuit, the resonant frequency of the circuit is $\omega_R = 1/\sqrt{LC}$, which is independent of the resistance, i.e., loss independent. Therefore, it is reasonable to adopt lossless assumption for the analysis. Furthermore, we also compared the resonance frequencies numerically predicted under lossy and lossless conditions. The following takes the odd-order MP resonances when $\beta = 0$ as an example to compare the differences in the predicted MP resonance frequencies under lossy and lossless conditions. Ag is chosen as the metal material and $SiO_2$ as the dielectric spacer. The optical constants of the material and the equivalent current depth are kept consistent with those selected in Section 3.1. With parameter $b$ held constant ($b$ = 0.05 μm), resonance frequencies are compared between the two cases by varying $h$. The results are shown in **Fig. 6**. The results show that there is almost no difference in the predicted MP resonance frequency under lossy conditions and lossless conditions. This confirmed the validity of lossless assumption. Hence the lossless approximation is adopted in the following discussions.

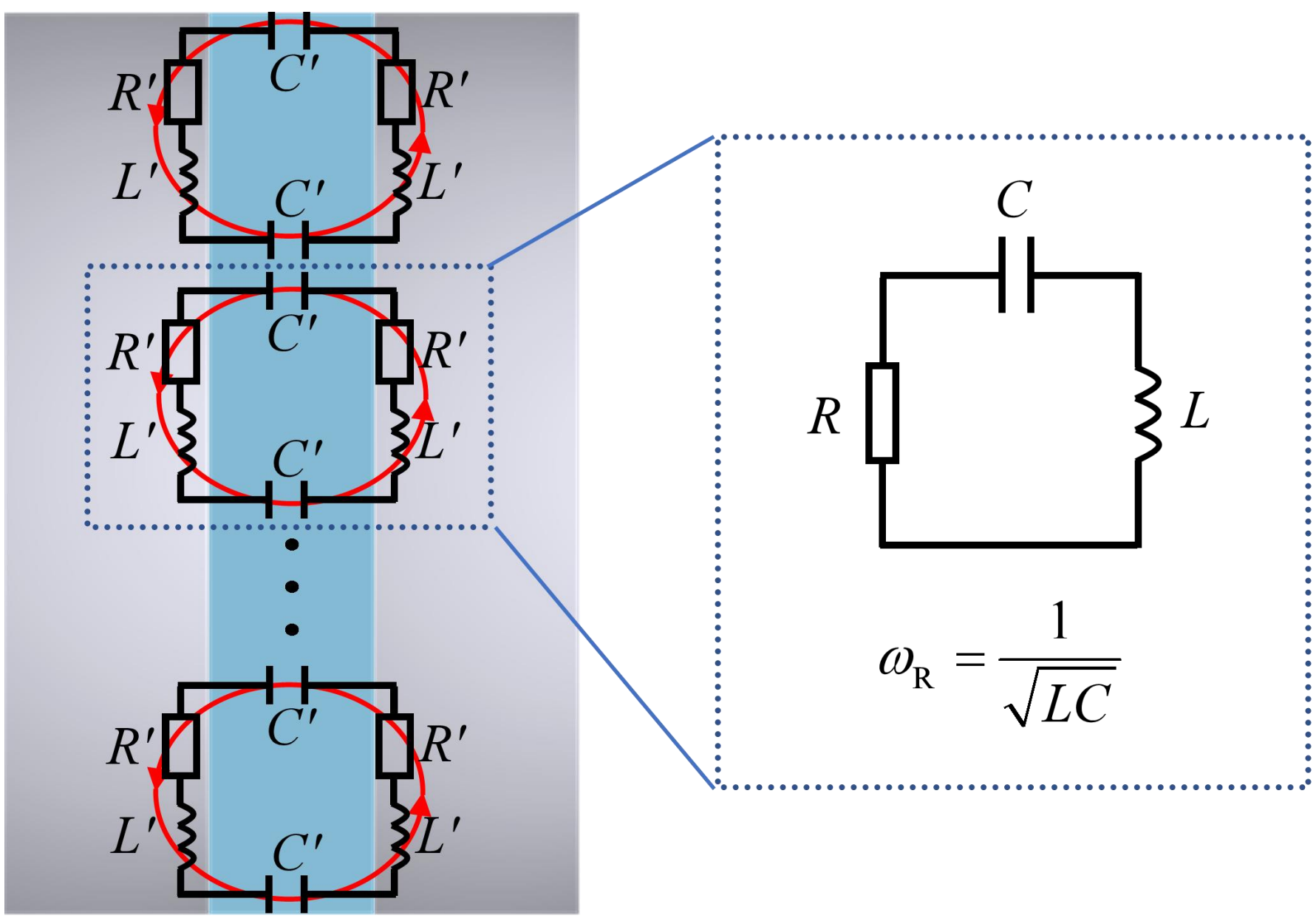

**Fig. 5** Schematic on analysis of the influence of loss on the resonance frequency of MP*m*.

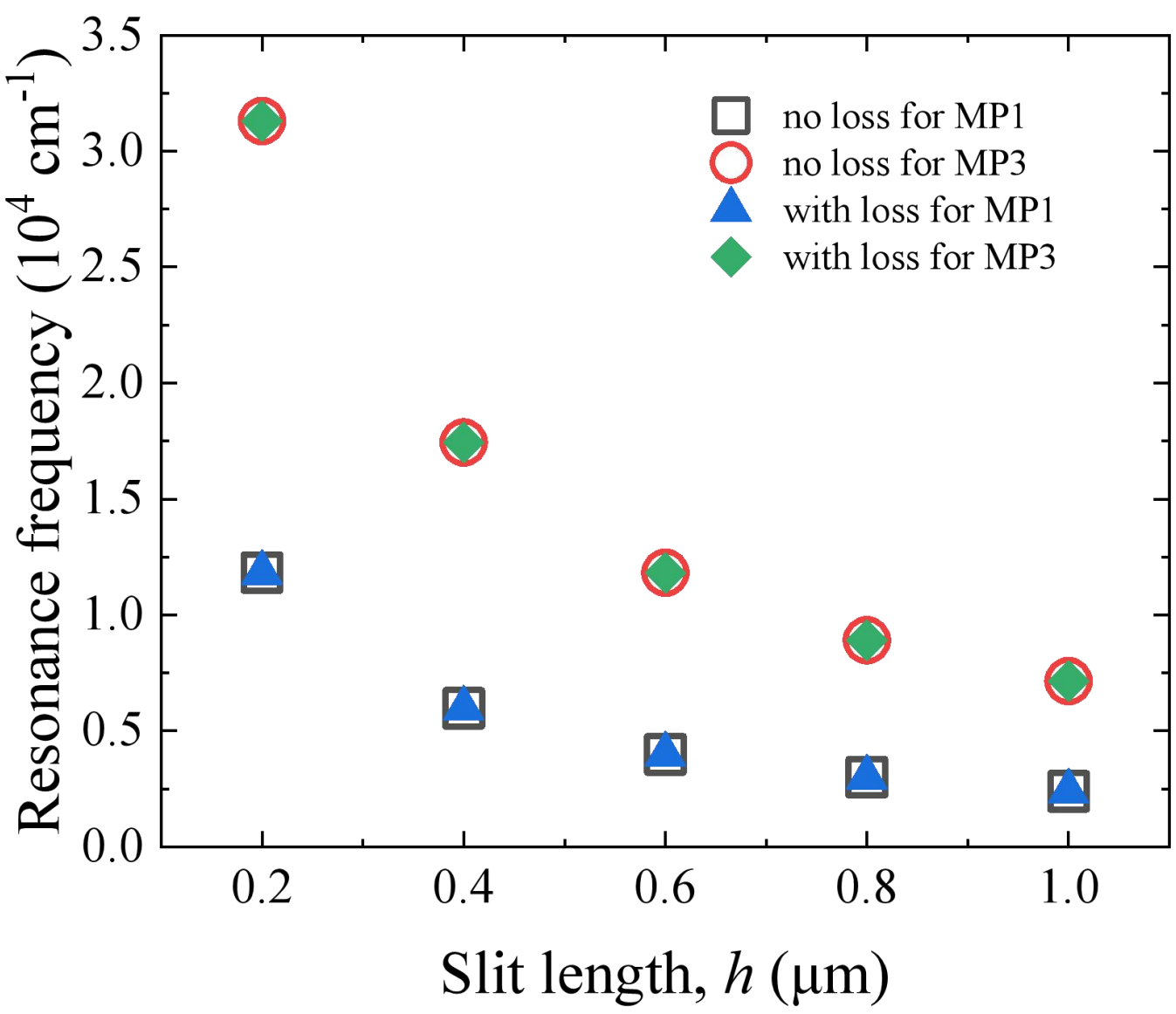

**Fig. 6** Comparison of the predicted MP resonance frequencies under lossy and lossless conditions.

For the horizontal MIM structure, as shown in **Fig. 4** (b), it is requisite to distinguish two cases of the incident light, namely normal incidence and oblique incidence. When the incident light is normally incident, only odd-order MP resonances can be excited within the structure. At this moment, it is necessary to take $\beta = 0$, that is, to adopt a completely symmetrical partition, and then Eq. (23)degenerates to

$$\begin{aligned}\tilde{Z}_{\text{tot}} &= \tilde{Z}_{\text{in, open}}\Big|_{z=-\frac{1}{2}h} + \tilde{Z}_{\text{in, open}}\Big|_{z=-\frac{1}{2}h} \\ &= -\mathrm{j}\sqrt{\frac{L_0(\omega)}{C_0(\omega)}}\frac{2}{\tan\left(\frac{1}{2}\omega h\sqrt{L_0(\omega)C_0(\omega)}\right)}\end{aligned} \tag{27}$$

The resonance condition for the case of normal incidence can be obtained as

$$\omega_{\text{MP}m} h\sqrt{L_0(\omega_{\text{MP}m})C_0(\omega_{\text{MP}m})} = m\pi, \quad m = 2n-1\ (n = 1, 2, \ldots) \tag{28}$$

In the case of oblique incidence, the resonance condition is identical to that of the vertical MIM structure. When employing the LCM to predict the fundamental-order MP resonance, as illustrated in **Fig. 1**(a), the resonance condition can be expressed as $\omega_{\text{MP1}} = 1/\sqrt{LC}$, where $L = \rho_{L,\text{m}}(\omega)\frac{h}{\delta(\omega)l} + 0.5\frac{\mu_0 bh}{l}$, $C = c_2\rho_{C,\text{i}}(\omega)\frac{hl}{b}$, $c_2$ is a numerical factor between 0.2 and 0.3 [23]. By comparison, it can be concluded that $L = hL_0/2$, $C = c_2 hC_0$. While $m$ =1, Eq. (26) can be written as $\omega_{\text{MP1}} = \pi/\sqrt{h^2 L_0 C_0}$. It can be further concluded that $c_2 = 2/\pi^2$, which is between 0.2 and 0.3. It is demonstrated that there is a connection between the LCM and DCM. Furthermore, by comparing the MP resonance condition derived from the DCM (Eq. (26)) with that of the GP mode (that is, by interpreting Eq. (26) within the framework of the Fabry-Perot resonator model), the phase constant $\gamma = \omega\sqrt{L_0(\omega)C_0(\omega)}$ and the reflection phase $\phi = 0$ can be extracted from Eq. (26). The results are consistent with those reported in reference [41]. This confirms the equivalence between the MP mode and the GP mode for this simple case.

For the metal groove, the initial end of the distributed circuit is located at the bottom of the groove, and a substrate impedance is required to be connected in series at the initial end, as depicted in **Fig. 4** (c). Assume that the impedance of the substrate $\tilde{Z}_{\text{sub}} = R_{\text{sub}} + \text{j}\omega L_{\text{sub}}$ (note: $\tilde{Z}_{\text{sub}}$ is a lumped impedance, and its value can be obtained by Eq. (8)). Similarly, assuming the real part can be disregarded, then the total impedance is determined as

$$\begin{aligned} \tilde{Z}_{\text{tot}} &= \tilde{Z}_{\text{sub}} + \tilde{Z}_{\text{in, open}} \big|_{z=-h} \\ &= \text{j}\omega L_{\text{sub}}(\omega) - \text{j}\sqrt{\frac{L_0(\omega)}{C_0(\omega)}}\cot(\omega h\sqrt{L_0(\omega)C_0(\omega)}) \\ &= \text{j}\sqrt{\frac{L_0(\omega)}{C_0(\omega)}}\left[k\xi(\omega) - \cot\xi(\omega)\right] \end{aligned} \tag{29}$$

where, $\xi(\omega) = \omega h\sqrt{L_0(\omega)C_0(\omega)}$ , $k = \dfrac{L_{\text{sub}}(\omega)}{hL_0(\omega)} \approx \dfrac{b}{2h}$ (If the substrate material is identical to that of the groove metal wall and the mutual inductance component in $L_0$ is neglected, $k$ can be approximated as a parameter associated with the geometric structure). The resonance condition can be written as

$$\cot\xi_m(\omega_{\text{MP}m}) = k\xi_m(\omega_{\text{MP}m}) \tag{30}$$

This is a transcendental equation that requires numerical methods for solving. $\xi_m$ is the root of the transcendental equation ( $\cot x = kx$ ) in the $m$-th period ($m\pi - \pi$, $m\pi$), as depicted in **Fig. 7**. After obtaining $\xi_m$ , the resonance frequency of the $m$-th order MP can be calculated as follows.

$$\omega_{\text{MP}m} = \frac{\xi_m(\omega_{\text{MP}m})}{h\sqrt{L_0(\omega_{\text{MP}m})C_0(\omega_{\text{MP}m})}} \tag{31}$$

It is noted that for extremely deep grooves ( $h \gg b$ ), $k \approx 0$ , Eq. (30) yields $\xi_m \approx m\pi - \pi/2$ , then the resonance condition (Eq. (31)) is simplified to $\omega_{\text{MP}m}h\sqrt{L_0(\omega_{\text{MP}m})C_0(\omega_{\text{MP}m})} = m\pi - \pi/2$ . In the view of GPs, the phase constant remains as $\gamma = \omega\sqrt{L_0(\omega)C_0(\omega)}$ , while the reflection phase becomes $\phi = \pi/2$ . However, in general case, the reflection phase is hard to be determined.

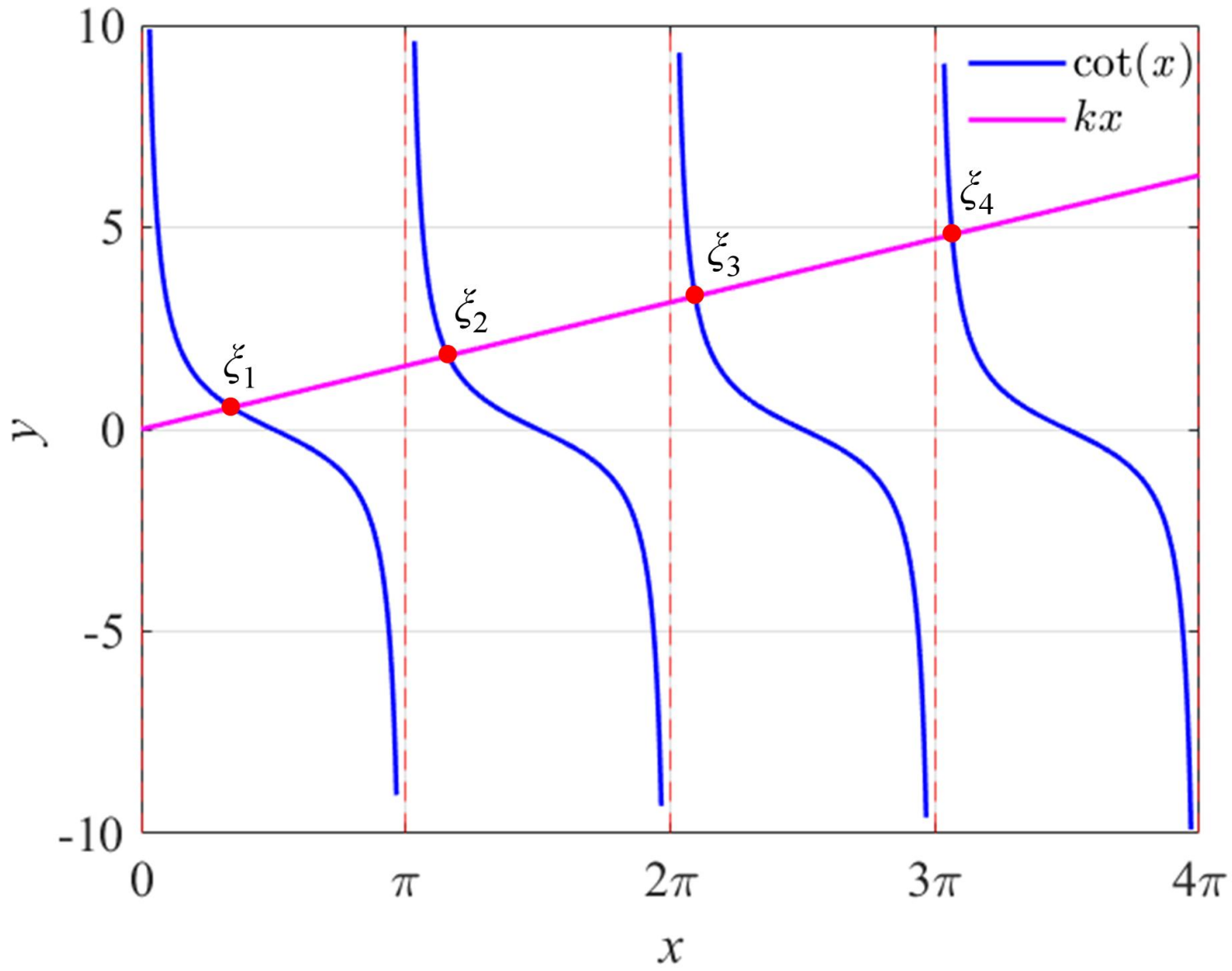

**Fig. 7** Plot of $y = \cot(x)$ and $y = kx$ ($k$ is approximately a constant associated with the geometric structure, and its value is set to 0.5 in the figure for illustrative purposes).

For the case where a thin dielectric layer is overlaid on a metal groove, the thin dielectric layer can be regarded as a load connected at the terminal of the circuit, as shown in **Fig. 4** (d). The load impedance $\tilde{Z}_{\mathrm{L}}$ can be obtained by Eq. (8). At this time, the input impedance is no longer a pure imaginary number, and the total impedance is

$$\begin{aligned}\tilde{Z}_{\mathrm{tot}} &= \tilde{Z}_{\mathrm{sub}} + \tilde{Z}_{\mathrm{in,\,load}}\big|_{z=-h} \\ &= \mathrm{j}\omega L_{\mathrm{sub}}(\omega) + \sqrt{\frac{L_0(\omega)}{C_0(\omega)}}\,\frac{\tilde{Z}_{\mathrm{L}}(\omega) + \mathrm{j}\sqrt{\frac{L_0(\omega)}{C_0(\omega)}}\tan\left(\omega\sqrt{L_0(\omega)C_0(\omega)}h\right)}{\sqrt{\frac{L_0(\omega)}{C_0(\omega)}} + \mathrm{j}\tilde{Z}_{\mathrm{L}}(\omega)\tan\left(\omega\sqrt{L_0(\omega)C_0(\omega)}h\right)}\end{aligned} \tag{32}$$

At this point, the imaginary part of the total impedance exhibits multiple zeros, each of which, in ascending order, corresponds to the resonance frequency of each order MPs.

$$\omega_{\mathrm{MP}m} = \text{the } m\text{-th zero of } \mathrm{Im}(\tilde{Z}_{\mathrm{tot}}) \tag{33}$$

In summary, **Table 1** compiles the resonance conditions of four typical MP structures derived from the DCM.

**Table 1** A summary of the resonance conditions of four typical MP structures derived from the DCM

| **MP structures** | **Graph of the MP structures** | **Resonance frequency of the *m*-th order MP** |
|---|---|---|
| Vertical MIM structure | | $\omega_{\mathrm{MP}m} h \sqrt{L_0(\omega_{\mathrm{MP}m}) C_0(\omega_{\mathrm{MP}m})} = m\pi, \quad m = 1, 2, \ldots$ |
| Horizontal MIM structure | | $\omega_{\mathrm{MP}m} h \sqrt{L_0(\omega_{\mathrm{MP}m}) C_0(\omega_{\mathrm{MP}m})} = m\pi,$<br>$m = 2n - 1 \ (n = 1, 2, \ldots)$<br>(Normal incidence)<br>$\omega_{\mathrm{MP}m} h \sqrt{L_0(\omega_{\mathrm{MP}m}) C_0(\omega_{\mathrm{MP}m})} = m\pi,$<br>$m = 1, 2, \ldots$<br>(Oblique incidence) |
| Metal groove | | $\cot \xi_m(\omega_{\mathrm{MP}m}) = k \xi_m(\omega_{\mathrm{MP}m})$<br>( $\xi_m$ is the root of the transcendental equation( $\cot x = kx$ ) in the *m*-th period ($m\pi - \pi$, $m\pi$),<br>$k = \frac{L_{\mathrm{sub}}(\omega)}{hL_0(\omega)} \approx \frac{b}{2h}, \xi(\omega) = \omega h \sqrt{L_0(\omega) C_0(\omega)}$ )<br>For deep groove ( $h >> b$ ),<br>$\omega_{\mathrm{MP}m} h \sqrt{L_0(\omega_{\mathrm{MP}m}) C_0(\omega_{\mathrm{MP}m})} = m\pi - \pi / 2$ |
| Metal groove with a thin dielectric cover | | $\omega_{\mathrm{MP}m}$ = the *m*-th zero of $\mathrm{Im}(\tilde{Z}_{\mathrm{tot}})$<br>$\tilde{Z}_{\mathrm{tot}} = \mathrm{j}\omega L_{\mathrm{sub}}(\omega) + \sqrt{\frac{L_0(\omega)}{C_0(\omega)}} \frac{\tilde{Z}_{\mathrm{L}}(\omega) + \mathrm{j}\sqrt{\frac{L_0(\omega)}{C_0(\omega)}} \tan\left(\omega\sqrt{L_0(\omega)C_0(\omega)}h\right)}{\sqrt{\frac{L_0(\omega)}{C_0(\omega)}} + \mathrm{j}\tilde{Z}_{\mathrm{L}}(\omega) \tan\left(\omega\sqrt{L_0(\omega)C_0(\omega)}h\right)}$<br>(The zeros, in ascending order, correspond to the resonance frequencies of MPs of each order, respectively.) |

## 3. Results and discussion

In this section, the resonance conditions presented in Section 2.3 are validated. The RCWA method is employed to calculate the radiative properties, electromagnetic field and dimensionless current density distribution of the structure [45]. In this study, the Fourier terms are 181 orders for the gratings to ensure the solution accuracy. A comparison is made between the theoretical predictions of the DCM and the RCWA calculations. By analyzing the electromagnetic field and the dimensionless current density distribution within the structure at the resonant frequency, the justification for the construction of the distributed circuit is elucidated.

### 3.1 Grating with vertical or horizontal MIM structure

Consider a one-dimensional (1D) grating with vertical MIM structure as shown in **Fig. 8** (a). The grating is described by a period Λ, metal strip thickness *w*, slit length *h*, and slit width *b*. The grating is surrounded by air and only transverse magnetic (TM) wave is considered. Ag is chosen as the metal and the optical properties of Ag are obtained from the Drude model $\tilde{\varepsilon}(\omega) = \varepsilon_\infty - \omega_p{}^2 / (\omega^2 - \mathrm{j}\Gamma\omega)$ with the following parameters[22]: plasma frequency $\omega_p = 1.39\times10^{16}$ rad/s, scattering rate $\Gamma = 2.70\times10^{13}$ rad/s, and a high-frequency constant $\varepsilon_\infty = 3.4$. $SiO_2$ is chosen for the dielectric spacer. Here, the refractive index of $SiO_2$ is assumed to be constant as 1.45 (Within the wavelength range of 0.5 to 20 μm, the refractive index of $SiO_2$ varies little with wavelength).

The geometric effects on the resonance conditions are illustrated in **Fig. 8** (b) and (d) by individually changing *h* or *b* from the base parameters ($b$ = 0.05 μm, $h$ = 0.8 μm and Λ = 1.2 μm). The spectral-directional reflectance *R* is calculated by the RCWA method. The contour plots show $1-R$ at normal incidence. The dashed lines are the results predicted by the DCM. The red, green and blue dashed lines respectively represent the predicted results of DCM when $c_1$ takes 1, 0.8 and 0.67 in the equivalent current depth. The results indicate that the prediction performance is optimal when $c_1$ is set to 0.8. Finally, the equivalent current depth is taken as $\delta = 0.8\lambda/(2\pi\kappa)$, in which $\kappa$ is the extinction coefficient of Ag. The equivalent current depth serves as a free parameter in the DCM. Here an optimized value is adopted after a systematic comparison. In studies of the traditional LCM, this free parameter is consistently present, with its value differing across various investigations [22, 29]. It is shown that the derived MP resonance conditions can accurately predict various multi-order MP resonance modes, with the exception of the coupling region between MPs and SPPs where the theoretical predictions exhibit significant deviations. The resonance frequencies of SPPs are only dependent on the period Λ, so its resonance frequency is fixed at 8330 $\mathrm{cm^{-1}}$ and 16500 $\mathrm{cm^{-1}}$. When the resonance frequencies of MPs are close to those of SPPs, the resonance splits into two branches. One branch is shifted towards higher frequencies, while the other one is shifted towards lower frequencies.

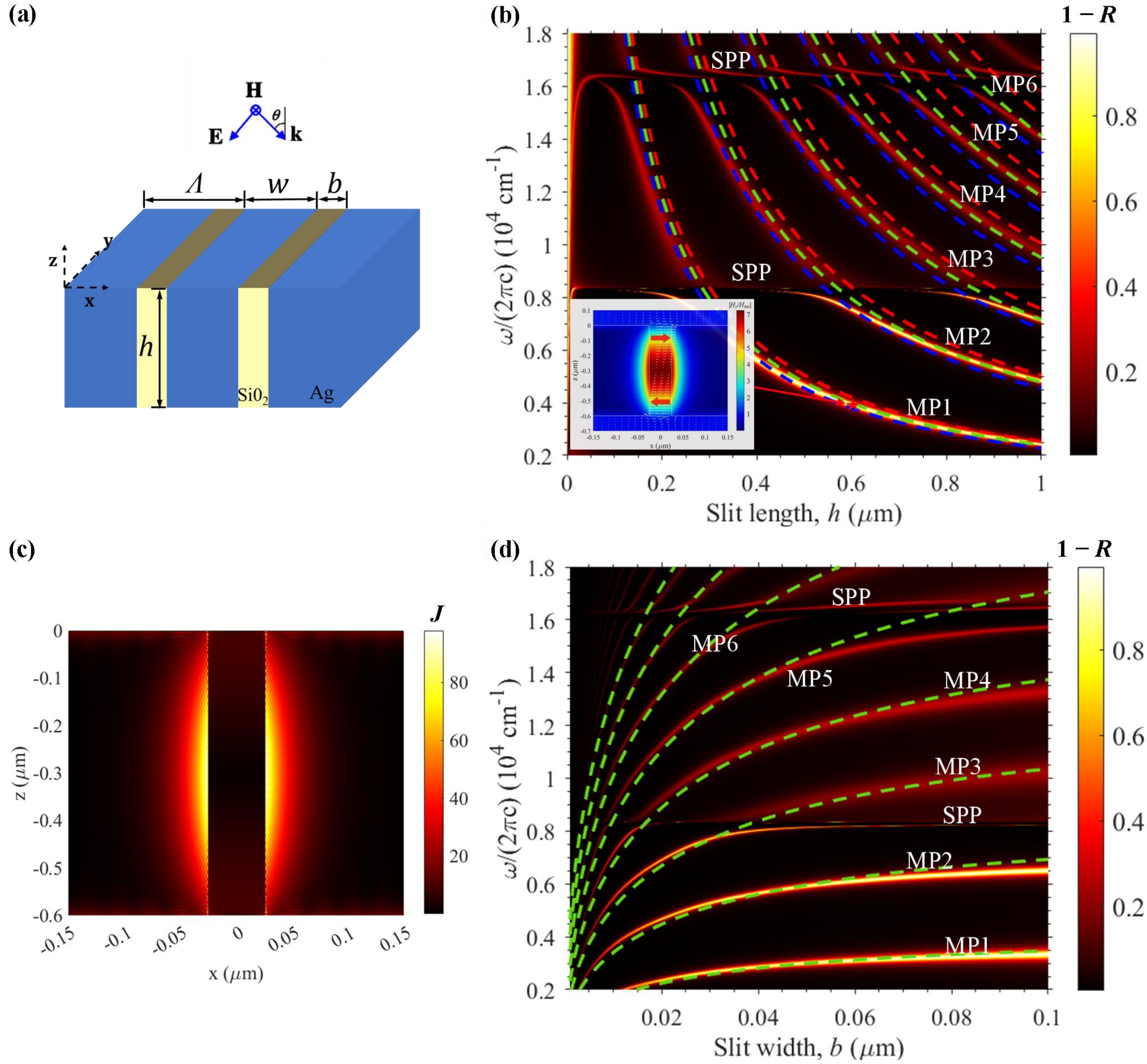

**Fig. 8** 1D grating with vertical MIM structure. **(a)** Schematic diagram. **(b, d)** Contour plots of $1-R$ as a function of wavenumber and varying grating height $h$ or slit width $b$ for a 1D grating with vertical MIM structure with $\theta = 0°$. The red, green and blue dashed lines respectively represent the predicted results of DCM when $c_1$ takes 1, 0.8 and 0.67 in the equivalent current depth. The inset in (b) is the electromagnetic field distribution in the grating for MP1 at 3922 cm$^{-1}$ with $\theta = 0°$. The color contours show the relative magnitude of the magnetic field, and the arrows indicate the direction and relative magnitude of the electric field vectors. **(c)** The dimensionless current density distribution in the 1D grating with vertical MIM structure for MP1 at 3922 cm$^{-1}$ with $\theta = 0°$. The dimensionless current density is defined as $J = \left| \mathrm{j}\omega\varepsilon_0\tilde{\varepsilon}\tilde{E} / \mathrm{j}\omega\varepsilon_0\tilde{E}_{\mathrm{inc}} \right| = \left| \tilde{\varepsilon}\tilde{E} / \tilde{E}_{\mathrm{inc}} \right|$.

When $b = 0.05$ μm, $h = 0.6$ μm and $\Lambda = 1.2$ μm, the resonance frequency of MP1 is 3922 cm$^{-1}$. The inset in **Fig. 8** (b) shows the electromagnetic field distribution, and **Fig. 8** (c) shows the dimensionless current density distribution within the structure at this resonance frequency. The dimensionless current density is defined as $J = \left| \mathrm{j}\omega\varepsilon_0\tilde{\varepsilon}\tilde{E} / \mathrm{j}\omega\varepsilon_0\tilde{E}_{\mathrm{inc}} \right| = \left| \tilde{\varepsilon}\tilde{E} / \tilde{E}_{\mathrm{inc}} \right|$. In the inset in **Fig. 8** (b), the color contours show the relative magnitude of the magnetic field, and the arrows indicate the direction and relative magnitude of the electric field vectors. Observe that the direction of the electric field is not uniform inside the slit. The direction of the electric field is opposite in the upper and lower parts of the slit. Therefore, the slit is divided into an upper and a lower part for processing. In **Fig. 8** (c), the color contours show the dimensionless current density magnitude. It can be seen that the position of the maximum current density on the metal wall is inside the slit, and that the current density on the slit wall gradually decreases from the middle to the two ends. In other words, the initial end of the distributed circuit should be set inside the slit. Therefore, we attempt to partition the slit into two unequal parts of a distributed circuit. The initial ends of the two parts are located at the same position within the slit, while the terminals are located at the two ends of the slit respectively. Both terminals are open circuits, and the impedances of the two parts are in series.

The reason why the slit is not vertically divided into two equal parts here is that we found that only resonant conditions for odd-order MP resonances can be obtained when the slit is symmetrically divided. When $b$ = 0.05 μm, $h$ = 0.6 μm and Λ = 1.2 μm, the resonance frequency of MP2 is 7510 $cm^{-1}$. **Fig. 9** (a) and (b) respectively show the electromagnetic field distribution within the structure at this resonance frequency when the incident angle $\theta$ is 0° and 30°. When MP2 is excited, there are two current loops of equal length within the slit. At $\theta$ = 0°, the electric fields in the upper and lower parts of a single current loop, which are in opposite directions, are symmetrically distributed around the center of the current loop, which is identical to the case of MP1. Nevertheless, at $\theta$ = 30°, the electric fields in the upper and lower parts of a single current loop, which are in opposite directions, are no longer symmetrically distributed around the center of the current loop. Consequently, an incomplete symmetrical division is necessary when establishing the DCM. For odd-order MP resonances, regardless of how the incident angle $\theta$ changes, the electric field within a single current loop is symmetrically distributed about the center of the current loop. This is the reason why an asymmetric division is required.

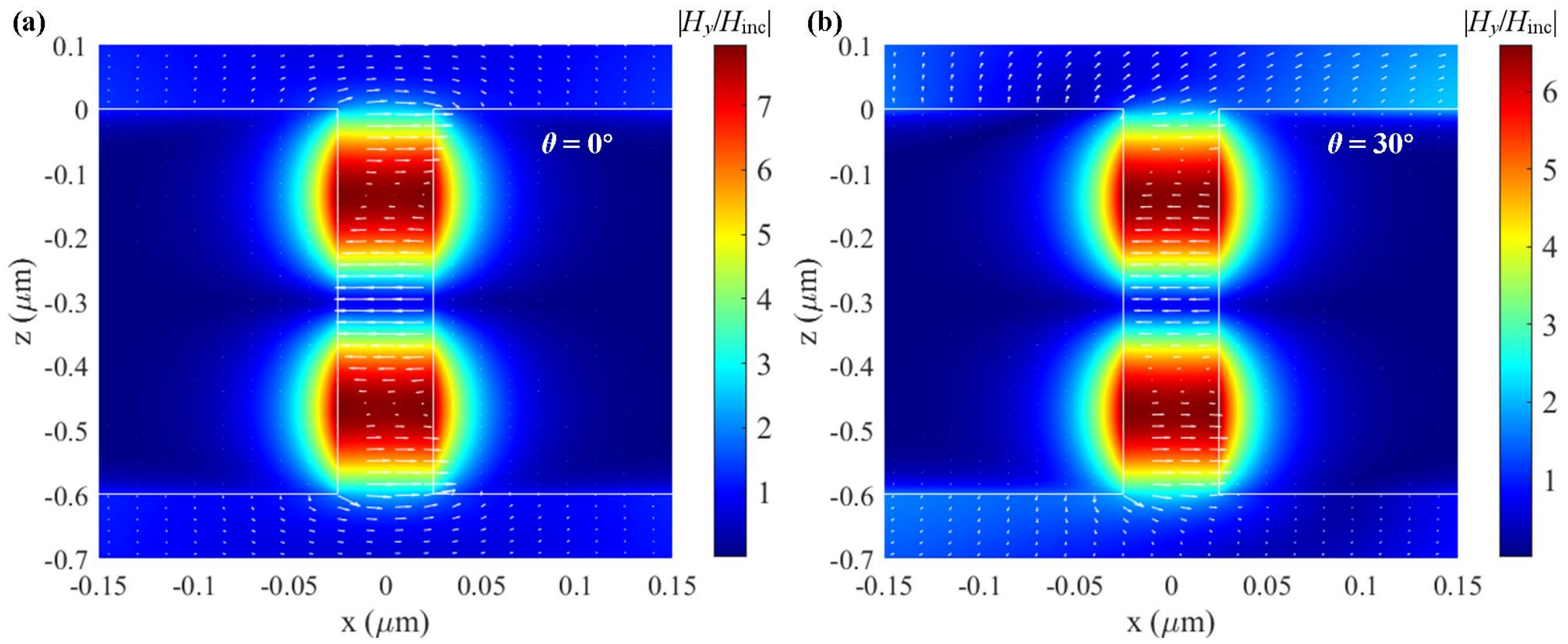

**Fig. 9** The electromagnetic field distribution for MP2 at 7510$cm^{-1}$ with **(a)** $\theta$ = 0° and **(b)** $\theta$ = 30°.

Next, consider a 1D grating with horizontal MIM structure as shown in **Fig. 10** (a). The geometric effects on the resonance conditions are illustrated in **Fig. 10** (b) and (c) by individually changing $h$ or $b$ from the base parameters ($b$ = 0.05 μm, $h$ = 0.8 μm, $w$ = 0.2 μm and Λ = 1.2 μm). The spectral-directional reflectance $R$ and transmittance $T$ are calculated by the RCWA method. The contour plots show $1-R-T$ at normal incidence. The dashed lines are the results predicted by the DCM (The equivalent current depth is taken as $\delta = 0.8\lambda/(2\pi\kappa)$ ). When the incidence is normal, only the odd-order MP resonance modes are present. Considering that the period remains constant, the SPP resonance frequency remains unchanged. At this time, however, the SPP at low frequencies is extremely weak and barely visible in figure. It can be observed that, in contrast to the 1D grating with vertical MIM structure, SPPs have a relatively minor effect on the resonance frequencies of MPs in the region of MP-SPP coupling under the horizontally placed conditions. Thus, the DCM exhibits better prediction.

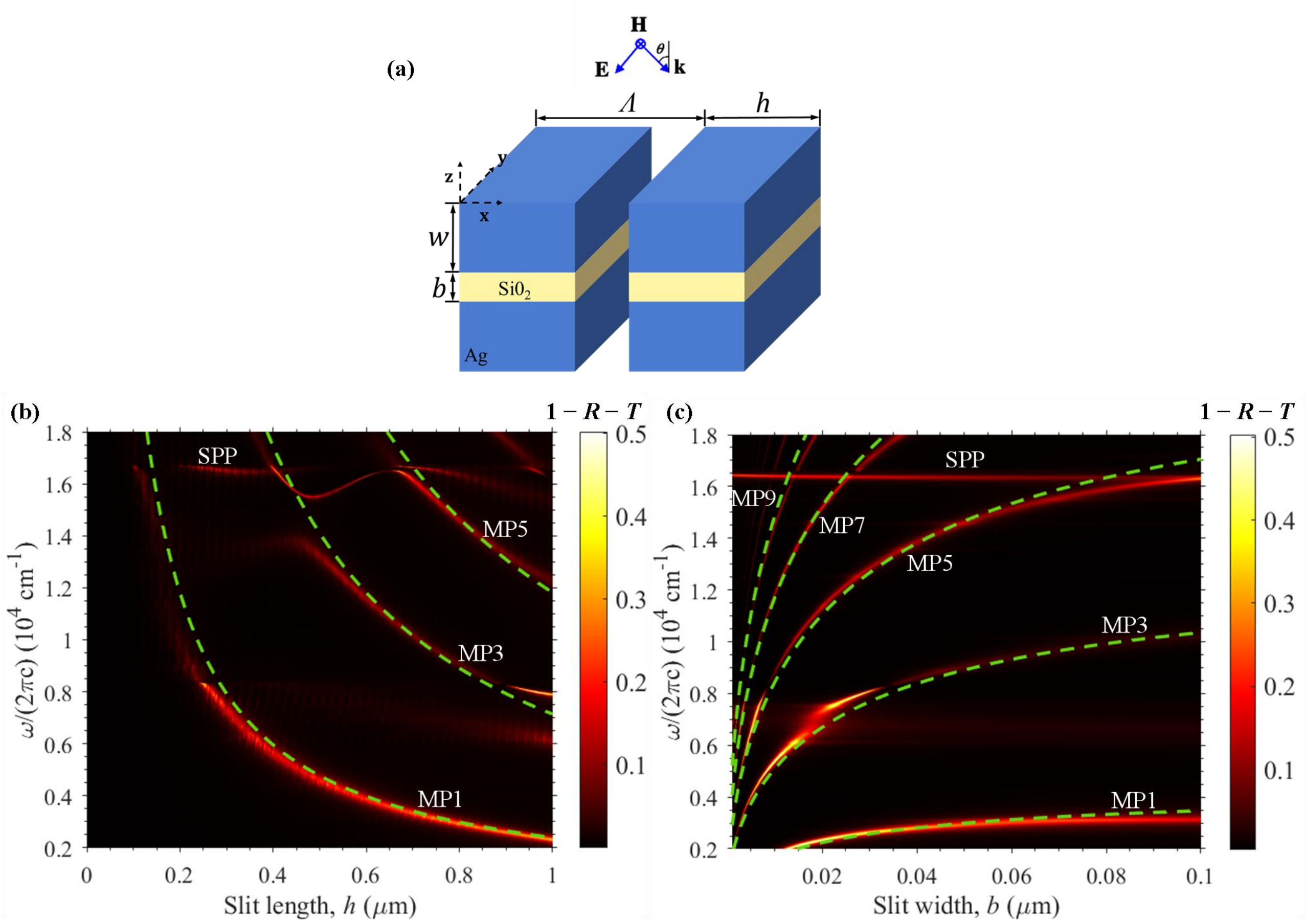

**Fig. 10** 1D grating with horizontal MIM structure. **(a)** Schematic diagram. **(b, c)** Contour plots of $1-R-T$ as a function of wavenumber and varying grating height h or slit width b for a 1D grating with horizontal MIM structure with $\theta = 0°$. The dashed lines are the resonance conditions predicted by the DCM.

## 3.2 Deep metal grating with substrate

Now, consider a 1D deep Ag grating with substrate. The substrate under the grating is made of the same material as the grating and is thick enough to be treated as opaque, as shown in **Fig. 11** (a). Under the base parameters ($b$ = 0.05 μm, $h$ = 0.6 μm and Λ = 0.5 μm), **Fig. 11** (b) and (d) show the geometric effects on the resonance conditions and the prediction of the DCM in this case. The red, green and blue dashed lines respectively represent the prediction results of DCM when $c_1$ takes the values of 1, 0.8 and 0.67. Similarly, the prediction effect is the best when $c_1$ is set to 0.8. The equivalent current depth is taken as $\delta = 0.8\lambda/(2\pi\kappa)$. It can be seen from the figure that the DCM has a larger error when the groove is wider for MP3. This is attributed to the fact that as the groove width increases, the resonance frequency of the MP3 increases and gradually approaches that of the

SPP.

When $b = 0.05$ μm, $h = 0.6$ μm, and $\Lambda = 0.5$ μm, the resonance frequency of MP1 is 2818 $cm^{-1}$. The inset in **Fig. 11** (b) shows the electromagnetic field distribution, and **Fig. 11** (c) shows the dimensionless current density distribution within the structure at this resonance frequency. In contrast to the case without a substrate, in the presence of a substrate, the directions of the electric field inside the groove are consistent. This way, the groove can be treated as an entire loop. Moreover, the current density on the two sidewalls is maximal at the bottom of the groove and gradually decreases from the bottom to the top of the groove walls. Hence, the initial end of the distributed circuit should be set at the bottom of the groove, and the terminal should be located at the top of the groove, which is an open circuit. The metal substrate is equivalent to a lumped impedance connected in series at the initial end of the distributed circuit. The lumped impedance of the metal substrate can be obtained by substituting $d$ with the groove width $b$ and $S$ with $\delta \times 1$ in Eq. (8).

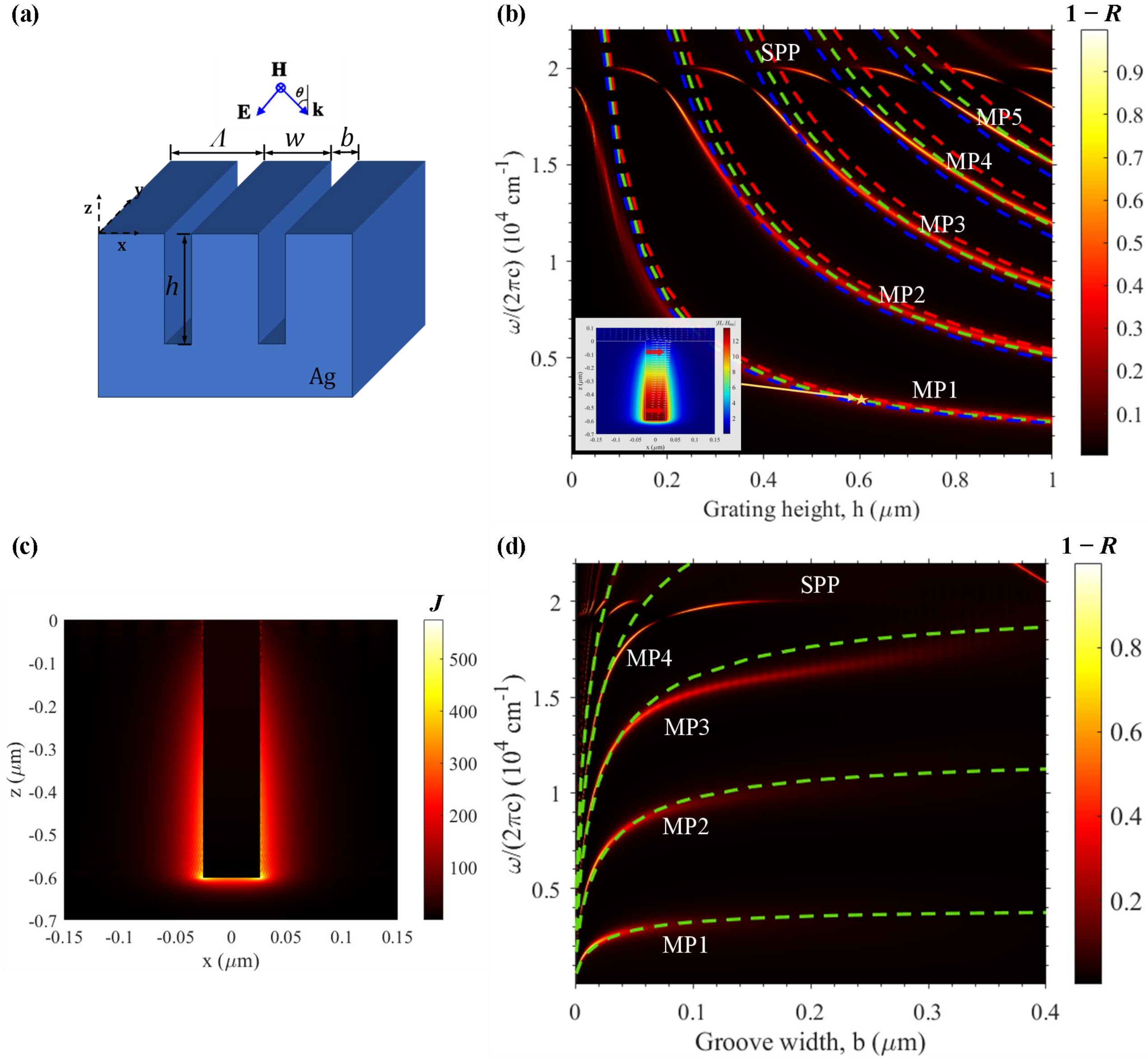

**Fig. 11** 1D deep Ag grating with substrate. **(a)** Schematic diagram. **(b, d)** Contour plots of 1−$R$ as a function of wavenumber and varying grating height $h$ or groove width $b$ for a 1D deep Ag grating with substrate with $\theta = 0°$.

The red, green and blue dashed lines respectively represent the predicted results of DCM when $c_1$ takes 1, 0.8 and 0.67. The inset in (b) is the electromagnetic field distribution in the grating for MP1 at 2818 $cm^{-1}$. **(c)** The dimensionless current density distribution in the 1D deep Ag grating with substrate for MP1 at 2818 $cm^{-1}$.

A ZnSe film is coated on the top of the 1D deep Ag grating with substrate as shown in **Fig. 12** (a). Within the wavelength range of 0.5 to 20 μm, ZnSe is a lossless transparent material with a refractive index that varies little with wavelength. Here, its refractive index is taken as 2.4. Adopting the same basic geometric parameters as those of the 1D metal grating mentioned previously, with the ZnSe film thickness $t$ = 15 nm, **Fig. 12** (b) and (d) show the geometric effects on the resonance conditions and the prediction of the DCM in this case (The equivalent current depth is taken as $\delta = 0.8\lambda/(2\pi\kappa)$). After the coating of the ZnSe film, the MP resonance frequency shifts towards lower frequencies overall, a trend that is well predicted by the DCM.

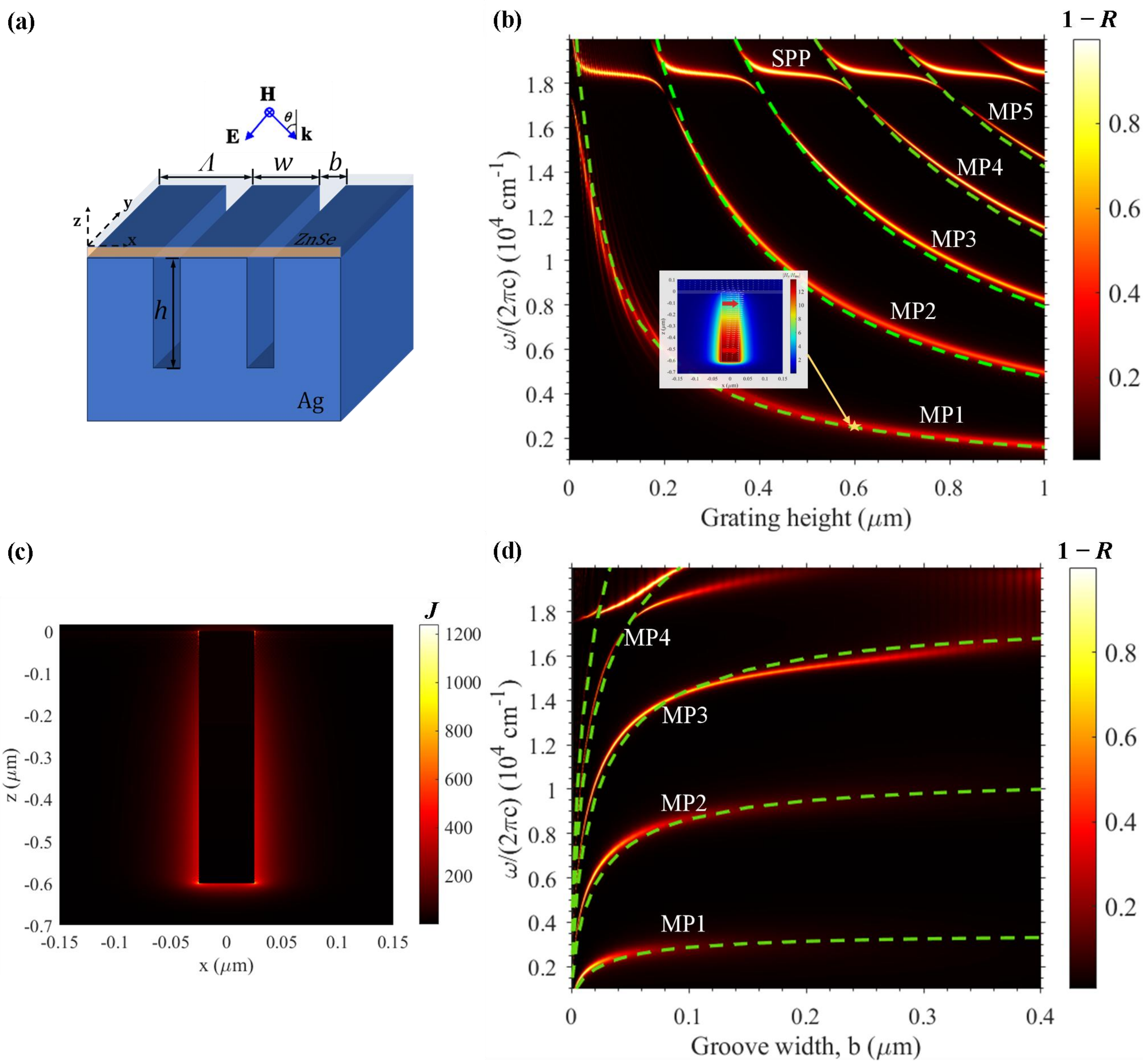

**Fig. 12** Thin-dielectric-layer-covered 1D deep Ag grating with substrate. **(a)** Schematic diagram. **(b, d)** Contour plots of $1-R$ as a function of wavenumber and varying grating height $h$ or groove width $b$ for a thin-dielectric-layer-covered 1D deep Ag grating with substrate with $\theta$ = 0°. The dashed lines are the resonance conditions predicted by the DCM. The inset in (b) is the electromagnetic field distribution in the grating for MP1 at 2584 cm-

[1]. **(c)** The dimensionless current density distribution in the thin-dielectric-layer-covered 1D deep Ag grating with substrate for MP1 at 2584 $cm^{-1}$.

When $b$ = 0.05 μm, $h$ = 0.6 μm, and $\Lambda$ = 0.5 μm, the resonance frequency of MP1 is 2584 $cm^{-1}$. The inset in **Fig. 12** (b) shows the electromagnetic field distribution, and **Fig. 12** (c) shows the dimensionless current density distribution within the structure at this resonance frequency. The coverage of the ZnSe film does not modify the electromagnetic field and dimensionless current density distribution within the groove. Hence, the initial end of the distributed circuit remains set at the bottom of the groove and the terminal is located at the top of the groove. In this case, however, the terminal is treated as connected to a load. The load impedance can be obtained by substituting $d$ with the groove width $b$ and $S$ with $t \times 1$ in Eq. (8). Considering ZnSe as a lossless dielectric, the load impedance at this time is a pure capacitive impedance.

## 4. Conclusions

In summary, a DCM is proposed to describe and predict the multi-order MP resonances. Considering that the displacement current is present everywhere between the metal walls when the MP is excited, it is more reasonable to employ a DCM to describe and predict the MP resonances. The imaginary part of the total circuit impedance derived from the DCM has multiple zeros, enabling it to predict both the fundamental and high-order MP resonances with a unified circuit. The corresponding DCMs are constructed based on the electromagnetic field and the dimensionless current density distribution when MPs are excited in different MP structures. Comparison of the predictions with the RCWA simulations yields a good agreement. In traditional LCMs, two free parameters are typically involved: the charge inhomogeneity coefficient and the equivalent current depth. In contrast, the DCM contains only one free parameter – the equivalent current depth. In studies related to GP modes, phase shift $\phi$ must be individually fitted for each order of the GP modes. However, in the DCM, a single, unified equivalent current depth is sufficient to yield accurate predictions for all orders of the MP resonances. Nevertheless, the DCM currently still exhibits certain limitations. For example, it is unable to make accurate predictions in the coupling region of MPs and SPPs, thus requiring further advances. When the structural dimensions and material properties are known, the resonant frequencies of various orders of MPs can be rapidly predicted using our model. This offers convenience to engineers in designing metamaterials that meet specific requirements. This study deepens the understanding and facilitates the design of MP-based thermal radiation metamaterials.

## Acknowledgment

JMZ thanks the support by the National Natural Science Foundation of China (No. 51976045) and the Young Scientist Studio Foundation of Harbin Institute of Technology.